%% file: main.tex
\documentclass[lettersize,journal]{IEEEtran}
\usepackage{amsmath,amsfonts}
\usepackage{algorithm}
\usepackage{array}
\usepackage[caption=false,font=normalsize,labelfont=sf,textfont=sf]{subfig}
\usepackage{textcomp}
\usepackage{stfloats}
\usepackage{url}
\usepackage{verbatim}
\usepackage{graphicx}
\usepackage{subfig}
\usepackage{algpseudocode}
\usepackage{cite}
\usepackage{xcolor}
\providecommand{\reviewmarkup}{0}
\ifnum\reviewmarkup=0
  \colorlet{red}{black}
\fi
\usepackage{multirow}
\newcommand{\method}{HDA-MoE}
\definecolor{color1}{RGB}{78, 121, 167}   % #4E79A7
\definecolor{color2}{RGB}{242, 142, 43}   % #F28E2B
\definecolor{color3}{RGB}{225, 87, 89}    % #E15759
\definecolor{color4}{RGB}{89, 161, 79}    % #59A14F
\definecolor{color5}{RGB}{176, 122, 161}  % #B07AA1
\makeatletter
\long\def\@makecaption#1#2{%
\ifx\@captype\@IEEEtablestring%
\footnotesize\bgroup\par\centering\@IEEEtabletopskipstrut{\normalfont\footnotesize #1}\\{\normalfont\footnotesize #2}\par\addvspace{0.15\baselineskip}\egroup%
\vskip 1pt
\else
\vskip 1pt
\setbox\@tempboxa\hbox{\normalfont\footnotesize {#1.}\nobreakspace\nobreakspace #2}%
\ifdim \wd\@tempboxa >\hsize%
\setbox\@tempboxa\hbox{\normalfont\footnotesize {#1.}\nobreakspace\nobreakspace}%
\parbox[t]{\hsize}{\normalfont\footnotesize\noindent\unhbox\@tempboxa#2}%
\else%
\ifCLASSOPTIONconference \hbox to\hsize{\normalfont\footnotesize\hfil\box\@tempboxa\hfil}%
\else \hbox to\hsize{\normalfont\footnotesize\box\@tempboxa\hfil}%
\fi\fi\fi}
\makeatother
\begin{document}
\bstctlcite{IEEEtran:BSTcontrol}

\title{\method: Hybrid Parallelism and Dynamic, Adaptive Scheduling for Mixture-of-Experts with 3D Near-Memory Processing}
\author{
Haochen Huang$^{1,2}$,
Shuzhang Zhong$^{1,2,4,5}$,
Shengxuan Qiu$^{1,2,3}$,
Zhe Zhang$^{4,5}$,
Shuangchen Li$^{4,5}$,
Cong Li$^{2}$,
Dimin Niu$^{4,5}$,
Hongzhong Zheng$^{4,5}$,
Guangyu Sun$^{2}$,
Runsheng Wang$^{2,6,7}$,
Meng Li$^{1,2,7*}$

% \thanks{$^\dagger$ Equal contribution.}

\thanks{$^*$ Corresponding author: meng.li@pku.edu.cn}

\thanks{$^1$ Institute for Artificial Intelligence, Peking University, Beijing, China.}

\thanks{$^2$ School of Integrated Circuits, Peking University, Beijing, China.}

\thanks{$^3$ School of Electronics Engineering and Computer Science, Peking University, Beijing, China.}

\thanks{$^4$ Alibaba DAMO Academy, Beijing, China.}

\thanks{$^5$ DAMO Academy, Alibaba Group, Beijing Hupan Lab, Hangzhou.}

\thanks{$^6$ Institute of Electronic Design Automation, Peking University, Wuxi, China.}

\thanks{$^7$ Beijing Advanced Innovation Center for Integrated Circuits, Beijing, China.}
}



\maketitle
\input{docs/0-abstract}
\input{docs/1-introduction}
\input{docs/2-background}
\input{docs/3-motivation}
\input{docs/4-method}

\input{docs/5-experiment}
\input{docs/6-conclusion}

\section*{Acknowledgments}
\textcolor{red}{\noindent\textbf{Generative AI Disclosure.} The authors used ChatGPT to assist with language editing and clarity in the manuscript and response letter. No figures, images, technical contributions, experimental results, or theoretical developments were generated by AI; all such content was independently developed and verified by the authors.}

\bibliographystyle{IEEEtran}
\bibliography{reference}
\makeatletter
\def\@IEEEBIOskipN{0\baselineskip}
\makeatother
\newcommand{\bioimage}[1]{\includegraphics[width=1in,height=1.25in,clip,keepaspectratio]{#1}}

\begin{IEEEbiography}[{\bioimage{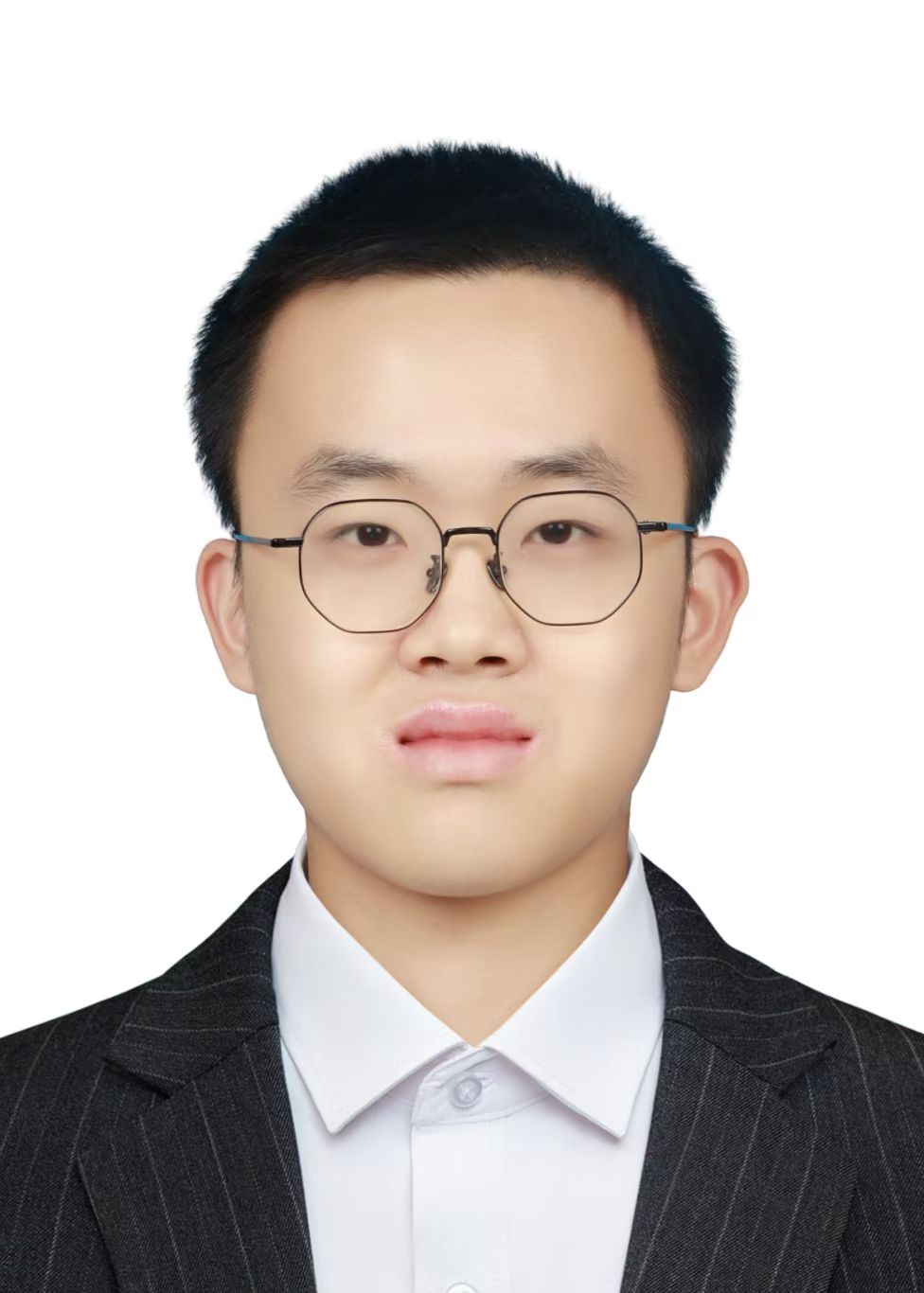}}]{Haochen Huang}
received the B.S. degree in applied physics from Peking University, Beijing, China, in 2026. He is pursuing the Ph.D. degree with the School of Integrated Circuits and the Institute for Artificial Intelligence, Peking University, Beijing, China. His current research interests include efficient AI systems and hardware--software co-design.
\end{IEEEbiography}

\begin{IEEEbiography}[{\bioimage{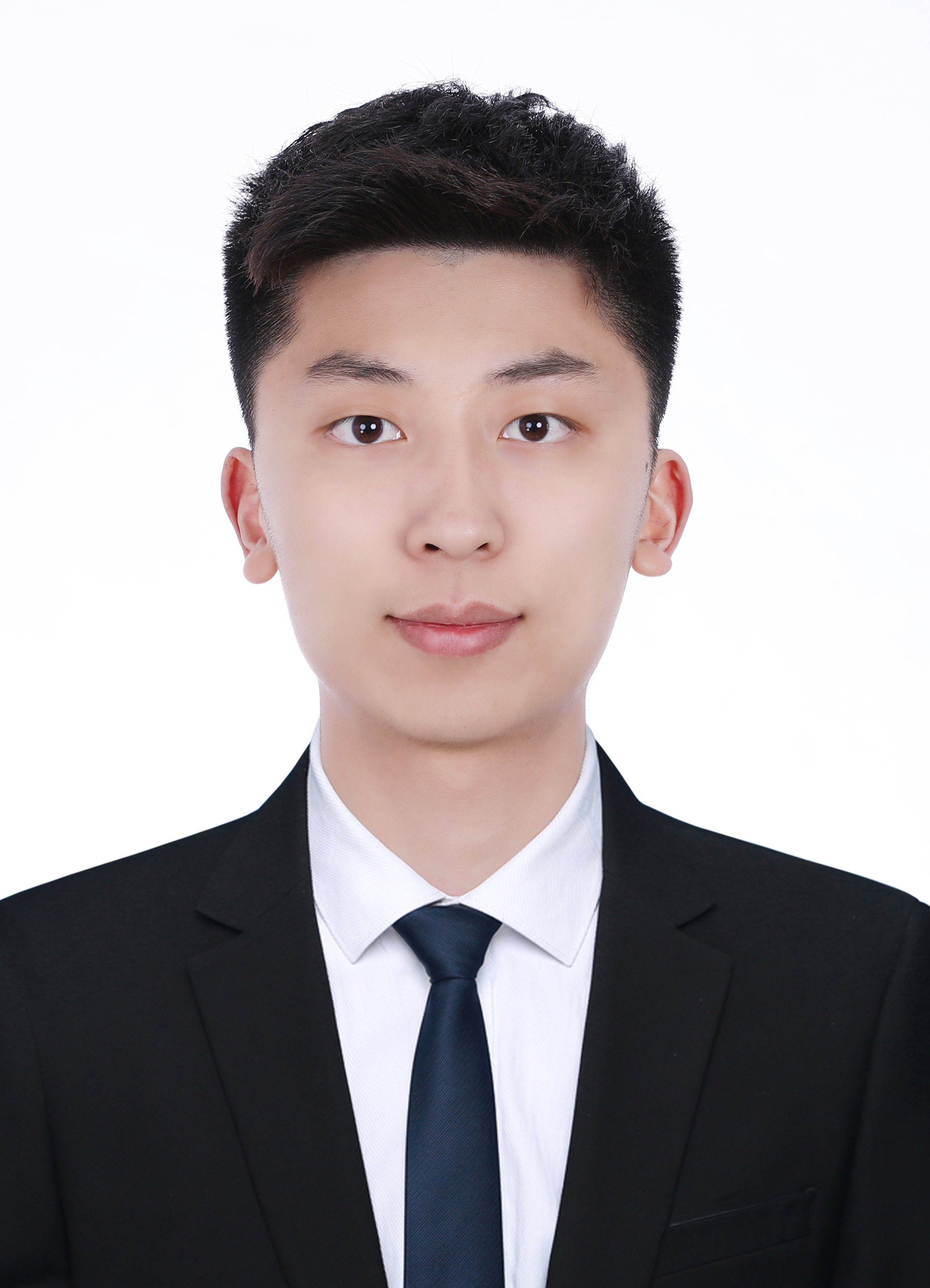}}]{Shuzhang Zhong}
received the B.S. degree in computer science and technology from Beihang University, Beijing, China, in 2023. He is currently pursuing the Ph.D. degree with the School of Integrated Circuits and the Institute for Artificial Intelligence, Peking University, Beijing, China. His current research interests include efficient LLM inference and agent system optimization.
\end{IEEEbiography}

\begin{IEEEbiography}[{\bioimage{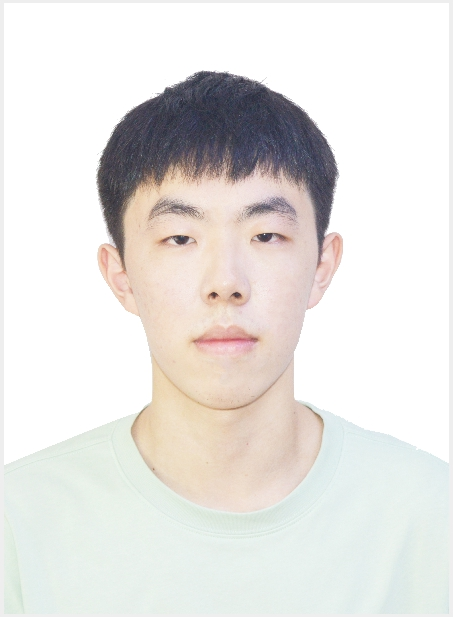}}]{Shengxuan Qiu}
is an undergraduate student at Peking University. His interests include efficient AI, large language model reasoning, and hardware-aware inference.
\end{IEEEbiography}

\begin{IEEEbiography}[{\bioimage{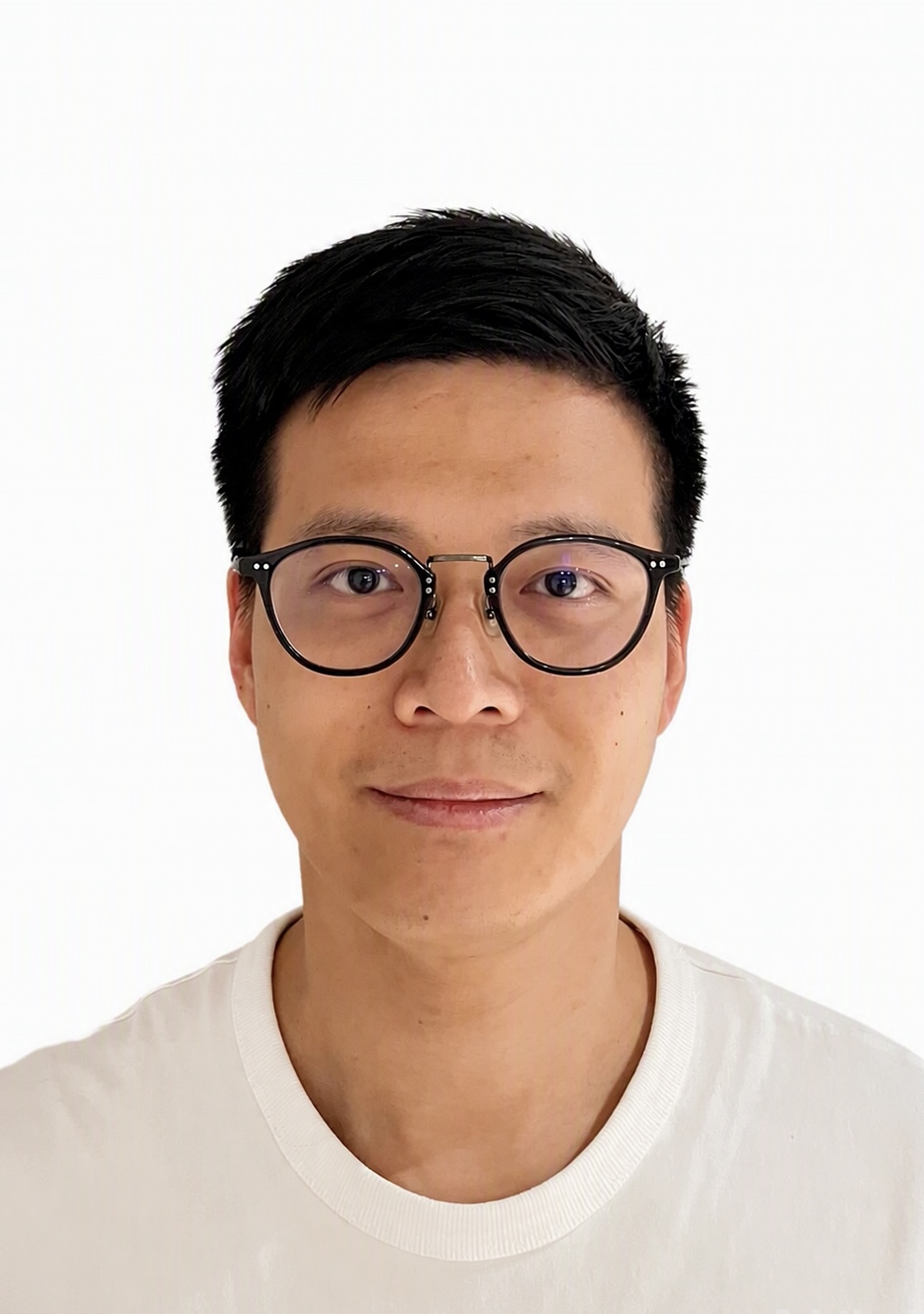}}]{Zhe Zhang}
received the Ph.D. degree in microelectronics from Peking University, Beijing, China, in 2020. He is currently a research scientist with the Computing Technology Laboratory, Alibaba DAMO Academy. His current research interests include computer architecture, domain-specific architecture, 3D-IC, and hardware--software co-optimization.
\end{IEEEbiography}

\begin{IEEEbiography}[{\bioimage{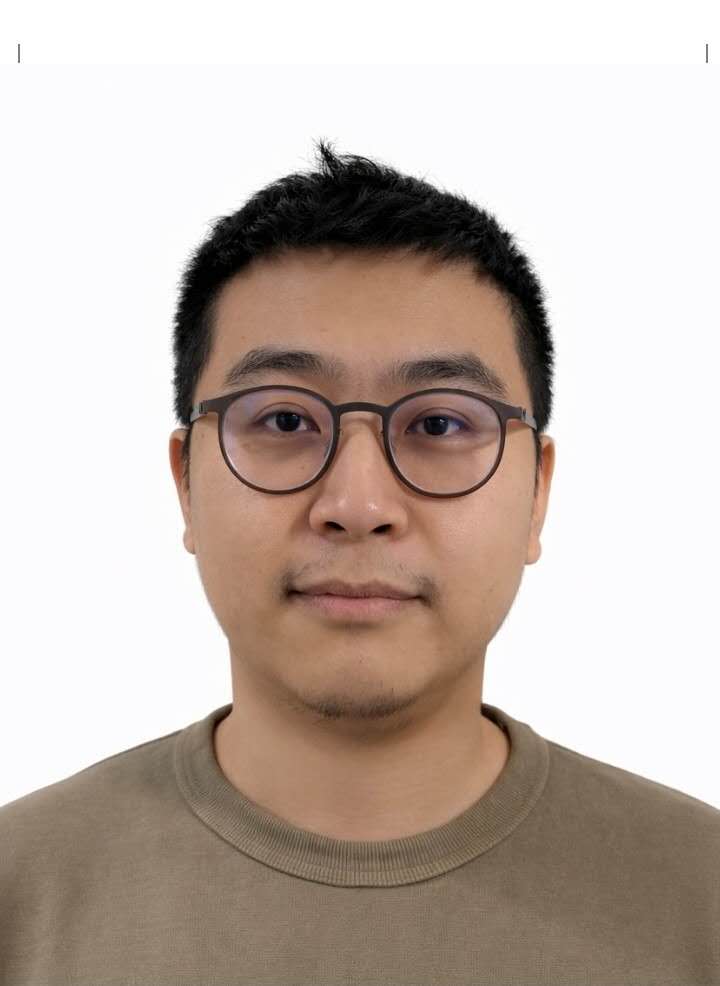}}]{Shuangchen Li}
received the B.S. and M.S. degrees from Tsinghua University, Beijing, China, in 2011 and 2014, respectively, and the Ph.D. degree in electrical and computer engineering from the University of California at Santa Barbara, Santa Barbara, CA, USA, in 2018. He is currently a research scientist with the Computing Technology Laboratory, Alibaba DAMO Academy, Sunnyvale, CA, USA. His interests include memory-related computer architecture, processing-in-memory architectures, emerging nonvolatile technologies, and deep learning accelerators.
\end{IEEEbiography}
% \vskip 8pt
\begin{IEEEbiography}[{\bioimage{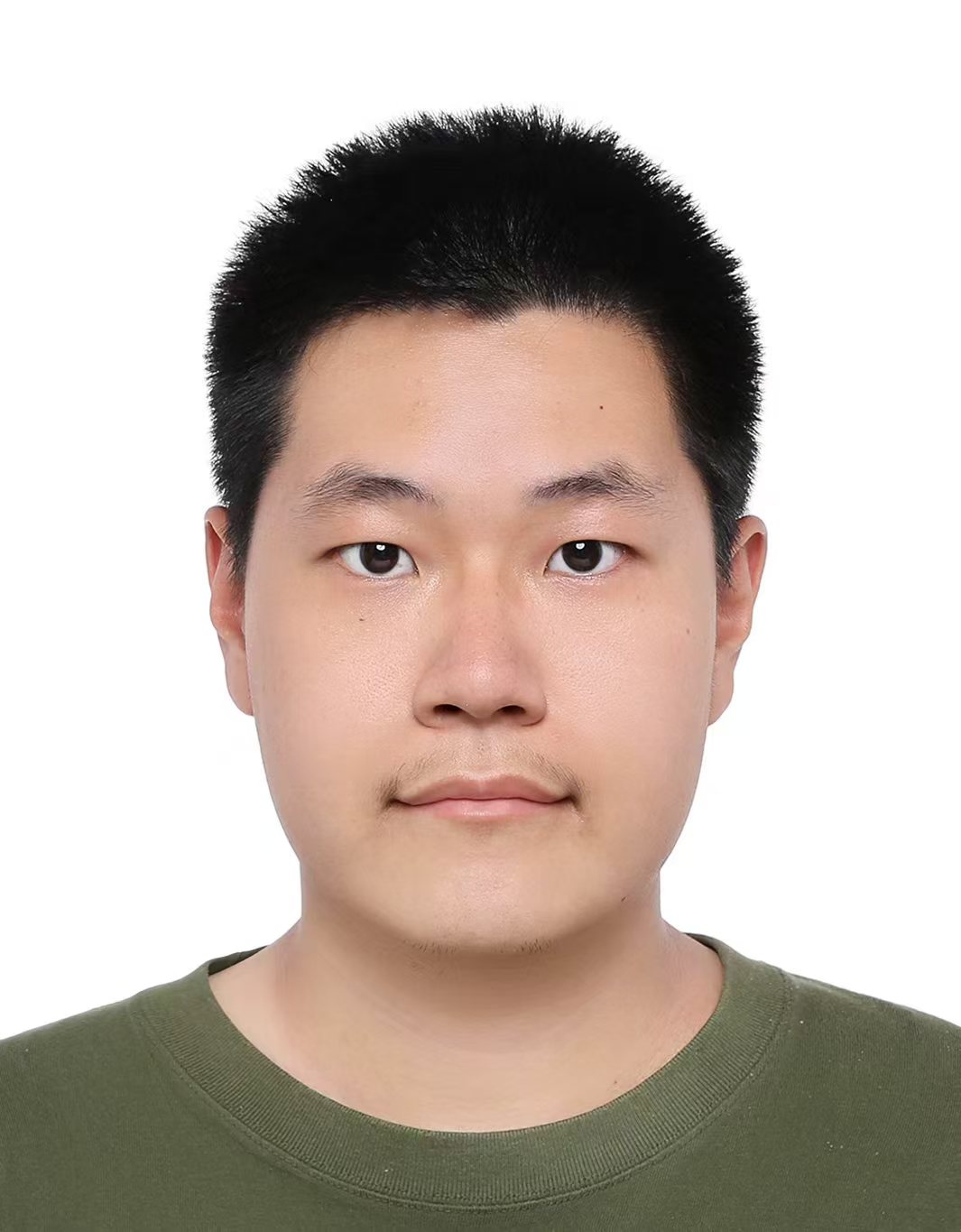}}]{Cong Li}
is a fourth-year Ph.D. candidate at Peking University, supervised by Prof. Guangyu Sun. His research interests include machine learning systems, domain-specific accelerators, and memory-centric computing architectures. Li received his bachelor's degree in computer science from Peking University. Contact him at leesou@pku.edu.cn.
\end{IEEEbiography}

\begin{IEEEbiography}[{\bioimage{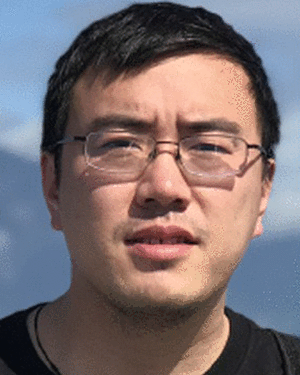}}]{Dimin Niu (Senior Member, IEEE)}
received the Ph.D. degree in computer science from Pennsylvania State University, University Park, PA, USA, in 2012. He was a staff memory architect with Memory Solutions Laboratory, Samsung Semiconductor Inc., San Jose, CA, USA. He is currently a research scientist with Computing Technology Lab, Alibaba DAMO Academy. His interests include computer architecture, memory architectures, storage systems, processing-in-memory, and domain-specific architectures.
\end{IEEEbiography}

\begin{IEEEbiography}[{\bioimage{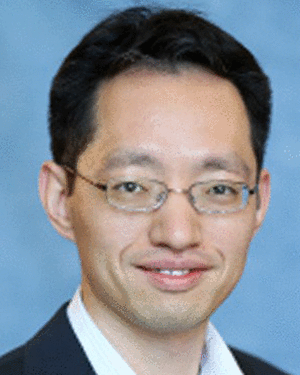}}]{Hongzhong Zheng (Member, IEEE)}
received the Ph.D. degree in computer engineering from the University of Illinois Chicago, Chicago, IL, USA. He was the director of Memory Solutions Lab, USA R\&D Center of Samsung Semiconductor, San Jose, CA, USA. He is currently a technical leader and research scientist with Computing Technology Lab, Alibaba DAMO Academy, Hangzhou, China. His interests include memory-system architecture, emerging memory technologies, processing-in-memory for machine learning, computer architecture, performance modeling, and energy-efficient computing. He is a member of ACM.
\end{IEEEbiography}

% \vskip 8pt
\begin{IEEEbiography}[{\bioimage{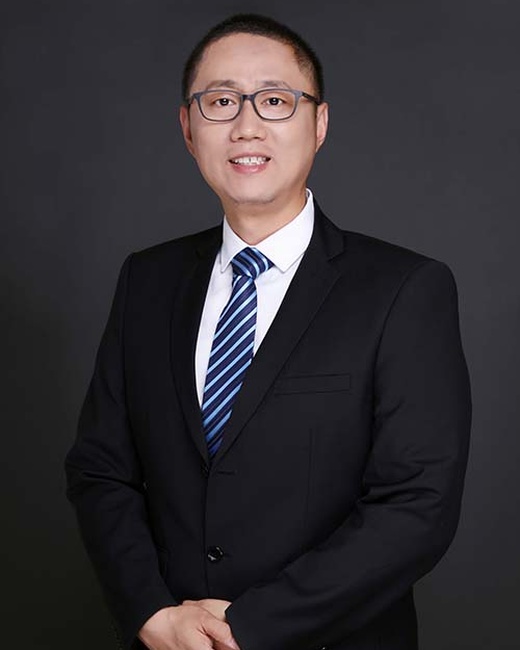}}]{Guangyu Sun (Senior Member, IEEE)}
is currently a Full Professor in the School of Integrated Circuits at Peking University. He received his B.S. and M.S. degrees from Tsinghua University, Beijing, in 2003 and 2006, respectively, and his Ph.D. degree from the Pennsylvania State University in 2011. His research interests include design and automation for computer architecture, cross-layer co-optimization, emerging memory technologies, etc. He has published 150+ journals and refereed conference papers on ISCA, MICRO, HPCA, DAC, IEEE TCAD, etc. His work has been recognized with the DAC Under-40 Innovators Award, CCF-IEEE CS Young Computer Scientists Award, Microsoft Research Asia Collaborative Research Award, CCF-Intel Young Faculty Researcher Program, and six best paper awards. He is an associate editor of IEEE TCAD.
\end{IEEEbiography}

% \vskip 8pt
\begin{IEEEbiography}[{\bioimage{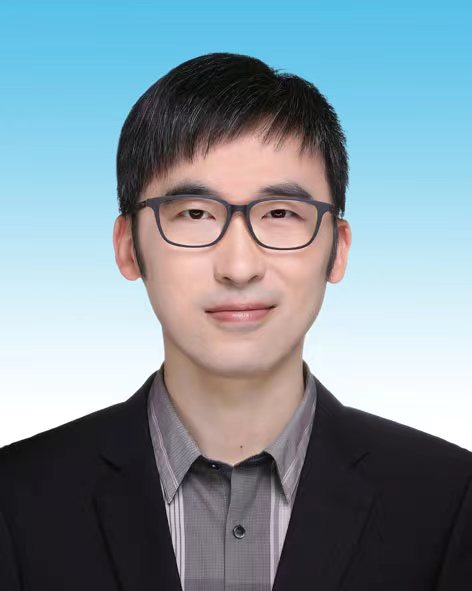}}]{Runsheng Wang (Senior Member, IEEE)}
received the B.S. and Ph.D. (highest honors) degrees from Peking University, Beijing, China, in 2005 and 2010, respectively. 

From November 2008 to August 2009, he was a Visiting Scholar with Purdue University, West Lafayette, IN, USA. He joined Peking University in 2010, where he is currently a Professor at the School of Integrated Circuits and is serving as the Associate Dean of the School of EECS. He has authored/coauthored 1 book, 4 book chapters, and about 200 scientific papers, including more than 40 papers published in \textit{International Electron Devices Meeting} (IEDM) and \textit{Symposium on VLSI Technology} (VLSI-T). He has been granted 19 US patents and 38 Chinese patents. His current research interests include nanoscale CMOS devices and reliability, design automation, and new-paradigm computing.

Dr. Wang was awarded the IEEE EDS Early Career Award by the IEEE Electron Device Society (EDS), National Distinguished Young Scholars by the National Natural Science Foundation of China (NSFC), Natural Science Award (First Prize) by the Ministry of Education (MOE) of China, and many other awards. He serves on the Editorial Board of \textit{IEEE TRANSACTIONS ON ELECTRON DEVICES}, and \textit{SCIENCE CHINA: Information Sciences}, and has served on the Technical Program Committee of many IEEE conferences, including \textit{IEDM}, \textit{IRPS}, etc.
\end{IEEEbiography}

% \vskip 8pt
\begin{IEEEbiography}[{\bioimage{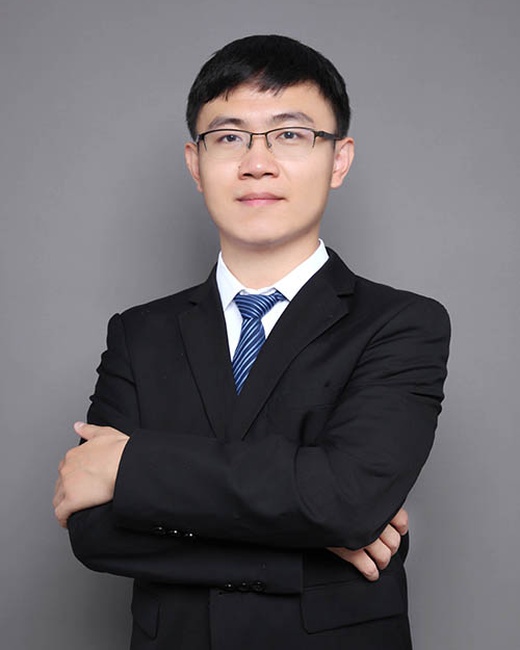}}]{Meng Li (Member, IEEE)}
received his Ph.D. degree in Electrical and Computer Engineering from the University of Texas at Austin in 2018 and is currently with the Institute for Artificial Intelligence and the School of Integrated Circuits, Peking University, Beijing, China. His research interests include efficient and secure multimodal AI acceleration hardware and algorithms. Before joining PKU, he was a staff research scientist and tech lead in the Facebook Reality Lab. He has published over 130 papers in premier conferences and journals with over 10,000 citations and several best paper awards/nominations. He is the recipient of the ACM SIGDA Outstanding New Faculty Award, CCF Integrated Circuit Early Career Award, Ant Group InTech Future Award, EDAA Outstanding Dissertation Award, and 1st Place in ACM Student Research Competition (Graduate Category), etc. 
\end{IEEEbiography}

\end{document}

%% file: docs/0-abstract.tex
\begin{abstract}
Mixture-of-Experts (MoE) architectures have become a key technique for scaling Large Language Models (LLMs), enabling high model capacity with reduced computational cost. However, this efficiency comes at the expense of increased memory capacity and bandwidth demands. Recent 3D Near-Memory Processing (NMP) architectures, which vertically integrate memory and compute through hybrid bonding, provide high internal bandwidth and energy efficiency, making them attractive for accelerating MoE inference. 
Nevertheless, the distributed memory and compute organization of NMP systems introduces new challenges for mapping MoE workloads. Existing parallelization strategies, such as Tensor Parallelism (TP) and Expert Parallelism (EP), suffer from either high communication costs or unbalanced computation utilization, leading to inferior efficiency. In addition, the dynamic routing behavior of MoE models further complicates efficient deployment.
To address these challenges, we present \method, a framework that optimizes MoE execution on NMP architectures through hybrid parallel deployment and runtime scheduling. \method~integrates an offline hybrid parallel mapping algorithm with an online dynamic and adaptive scheduling mechanism to reduce communication overhead while improving computation utilization. 
Experimental results show that \method~achieves a speedup of 1.1$\times$--3.4$\times$ over TP, 1.1$\times$--1.5$\times$ over EP, 1.1$\times$--3.7$\times$ over the Hybrid TP-EP compute-balanced baseline, and 1.1$\times$--1.3$\times$ over HD-MoE.
Source code is available at \url{https://github.com/PKU-SEC-Lab/HDA-MoE-TCAD26}.
\end{abstract}

\begin{IEEEkeywords}
Automated Deployment, Mixture-of-Experts, 3D Near-Memory Processing, Hardware-Aware Routing, NoC Simulation
\end{IEEEkeywords}

%% file: docs/1-introduction.tex
\section{Introduction}

\IEEEPARstart{R}{ecent} advances in Mixture-of-Experts (MoE) have made it a widely used architecture for scaling Large Language Models (LLMs) \cite{deepseekv3,jiang2024mixtral}. By activating only a small subset of experts for each token, MoE reduces computational cost while maintaining large model capacity. However, the sparse activation pattern introduces irregular memory accesses and often aggravates memory bottlenecks, especially on edge devices with limited bandwidth and small batch sizes.

% The recent emerging 3D Near-Memory Processing (NMP) architectures seem to be a promising solution for memory-bound problems \cite{memorywall,Near-memory-FPGAs}. 3D NMP vertically stacks DRAM dies directly on top of logic dies using high-bandwidth interconnects. In contrast to conventional von Neumann architectures, the vertical stacking of 3D NMP allows multiple memory banks to be accessed independently and in parallel, enabling fine-grained, high-throughput data access. This makes 3D NMP particularly suitable for MoE inference workloads.

Recent emerging 3D Near-Memory Processing (NMP) architectures provide a promising platform for such bandwidth-intensive workloads \cite{memorywall,Near-memory-FPGAs}. By vertically stacking DRAM on logic dies through high-bandwidth interconnects, NMP enables fine-grained parallel memory access and high internal bandwidth, making it well-suited for MoE inference.

% While MoE's bandwidth efficiency makes it suitable for 3D NMP deployment, the architectural shift from GPU-style shared memory to distributed NoC-based designs introduces new mapping challenges. The distributed nature of 3D NMP, with its bank-local memory organization, requires careful co-design of expert parallelism and communication routing strategies to maintain performance. As illustrated in Fig.~\ref{fig:intro}, current approaches employ either Tensor Parallelism \cite{TP} (TP) or Expert Parallelism \cite{lepikhin2020gshard} (EP): TP distributes each expert's parameter tensor across banks while EP assigns complete experts to different banks. This presents a fundamental trade-off: TP achieves better workload balance but incurs substantial all-reduce communication overhead, whereas EP minimizes communication but suffers from workload imbalance due to varying expert utilization. 

\begin{figure}[!tb]
    \centering
    \includegraphics[width=0.96\linewidth]{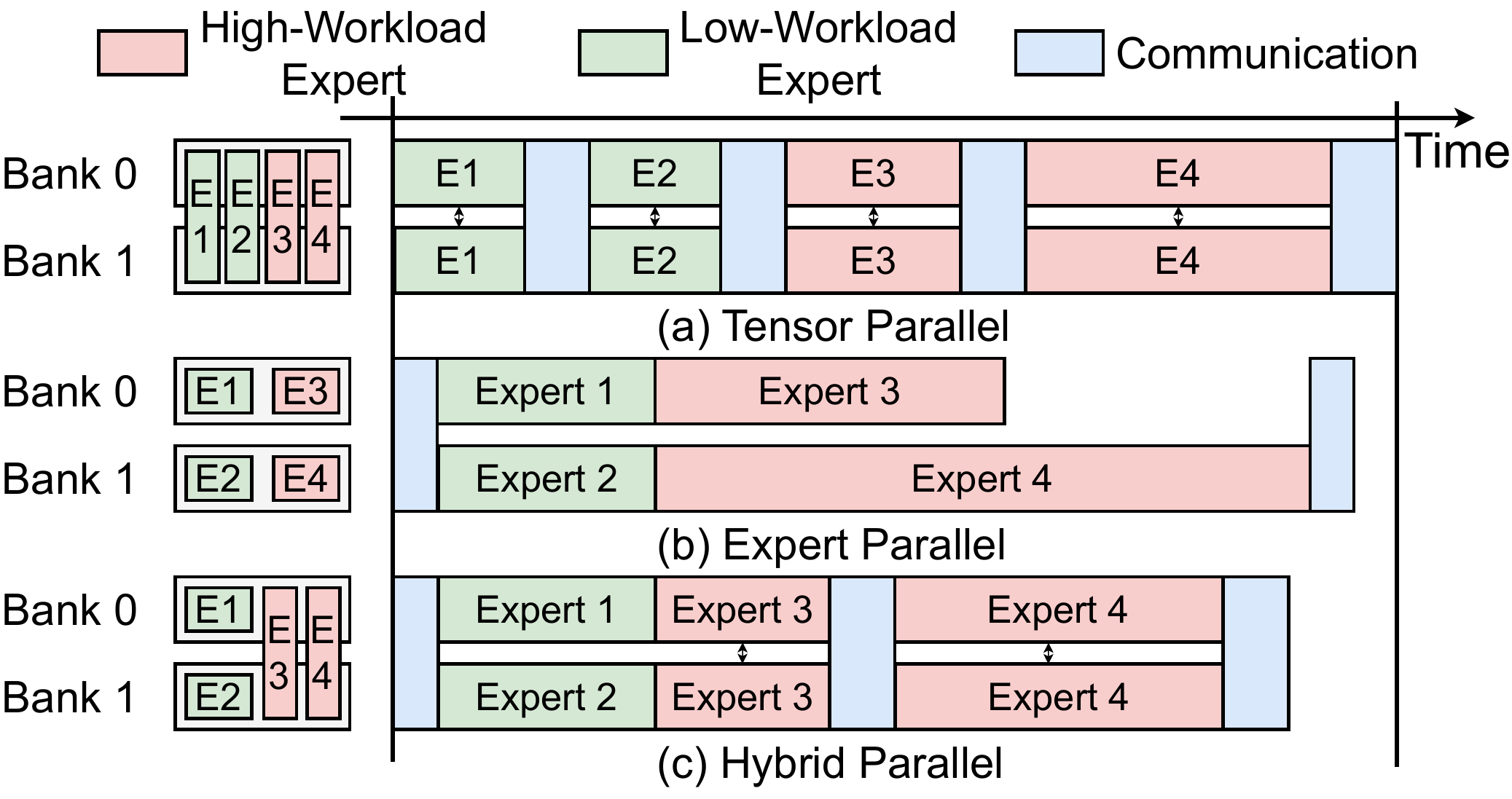}
    \caption{Comparison of expert placement and execution timelines under (a) tensor parallelism, (b) expert parallelism, and (c) hybrid parallelism.}
    \label{fig:intro}
\end{figure}

However, deploying MoE models on NMP systems introduces new challenges. Unlike GPUs with large shared memory, NMP architectures consist of distributed memory banks connected by an on-chip network (NoC), requiring careful coordination between expert placement and communication. As shown in Fig.~\ref{fig:intro}, Tensor Parallelism (TP) \cite{TP} partitions each expert across banks, improving load balance but incurring heavy all-reduce communication, whereas Expert Parallelism (EP) \cite{lepikhin2020gshard} places complete experts on different banks, reducing communication but suffering from workload imbalance.

% Previous works on GPU clusters have explored combining EP with replication of frequently activated experts to achieve both workload balance and low communication overhead. This method has been adopted by DeepSeek-AI to deploy its DeepSeek-R1 model \cite{deepseekv3}. However, this approach is impractical for 3D NMP due to its limited memory capacity. Furthermore, the dynamic and imbalanced nature of expert activation patterns significantly complicates mapping and scheduling decisions, requiring more sophisticated optimization strategies tailored to the constraints of 3D NMP architectures.

Large GPU clusters can combine EP with expert replication to mitigate imbalance, as in DeepSeek-R1 deployment \cite{deepseekv3}. However, replication is less practical for memory-constrained 3D NMP, and dynamic expert activation further makes static deployment insufficient.

% To address the challenge of dynamic expert activation, several studies focusing on offloading scenarios have investigated dynamic scheduling of experts \cite{AdapMoE, lin2024task, zhang2024daop, tang2024hobbit}. In these scenarios, experts are stored in secondary storage, with on-demand loading becoming the primary bottleneck. These studies demonstrate that MoE models often exhibit high activation similarity between adjacent layers, which can be exploited for prefetching to alleviate the on-demand loading overhead.

Recent offloading works study dynamic expert scheduling \cite{AdapMoE,zhang2024daop,tang2024hobbit}, where experts are loaded on demand and transfer latency dominates. They also observe temporal locality across adjacent layers, which can guide prefetching.

In addition, the contribution of different experts to the final output is often uneven. While a small subset of experts receives high routing scores and dominates the computation, many others have relatively low influence on the output~\cite{tang2024hobbit,AdapMoE,wang2025buddymoe,lu2024not}. This suggests that certain experts may be interchangeable without significantly affecting model accuracy, providing opportunities to guide expert selection toward more balanced and communication-efficient execution.

% In light of these challenges and opportunities, we propose \textbf{\method}, a hybrid parallelism and dynamic, adaptive scheduling framework designed for MoE inference on 3D NMP architectures. Reducing latency on distributed systems requires both \textbf{balancing computation utilization} and \textbf{minimizing communication cost}, while also addressing memory limitations. To achieve this, \method~ adopts a hybrid parallelism approach, as illustrated in Fig.~\ref{fig:intro}(c). Experts with low activation frequency are mapped using Expert Parallelism to minimize communication overhead, while high-frequency experts utilize Tensor Parallelism to maximize computational resource utilization. Additionally, \method~ incorporates online dynamic expert placement and hardware-aware gating strategy to mitigate the impact of the dynamic activation pattern.

Motivated by these challenges, we propose \textbf{\method}, a hybrid parallelism and dynamic, adaptive scheduling framework for MoE inference on 3D NMP. As shown in Fig.~\ref{fig:intro}(c), \method~uses EP for low-activation experts to reduce communication and TP-style partitioning for frequent experts to improve utilization. It further introduces online dynamic placement and hardware-aware adaptive gating to handle dynamic activation patterns.

Building upon HD-MoE~\cite{huang2025hd}, this journal version adds hardware-aware gating, analyzes low-impact expert substitutability, expands system modeling, and extends evaluation with conference-baseline comparison, latency breakdown, accuracy, scalability, and sensitivity studies. Code and evaluation artifacts are available at \url{https://github.com/PKU-SEC-Lab/HDA-MoE-TCAD26}.

Our key contributions are summarized as follows:
\begin{itemize}
    \item 
    % \textbf{Performance Analytical Model.} 
    % We develop a unified performance analysis framework applicable to diverse hardware configurations and parallelism strategies.
    We build a \textbf{unified performance model} to estimate MoE computation and communication cost and guide deployment optimization.
    \item \textbf{Automated Hybrid Parallelism.} 
    % We propose an efficient placement strategy searching method that combines TP and EP to optimize computation and communication overheads.
    We develop a placement framework that combines TP and EP to jointly optimize workload balance and communication overhead.
    \item \textbf{Dynamic Scheduling.} 
    % We introduce a dynamic expert placement strategy, which adjusts expert deployment in real-time based on the inference workload, ensuring optimal performance even in different inference scenarios.
    \textcolor{red}{We introduce runtime scheduling to reduce latency caused by short-term expert-activation fluctuations under a given routing pattern.}
    \item \textbf{Hardware-aware Gating.} \textcolor{red}{We introduce hardware-aware gating to reduce routing-induced computation and communication bottlenecks with limited model perturbation.}
    \item 
    % We conduct extensive experiments to validate our approach, demonstrating significant improvements in both TBT latency and speedup compared to baseline methods.
    Extensive experiments demonstrate clear improvements in TBT and speedup over strong baseline methods.
\end{itemize}

%% file: docs/2-background.tex
\section{Background}

\subsection{Mixture-of-Experts Models}

\begin{figure}
    \centering
    \includegraphics[width=0.95\linewidth]{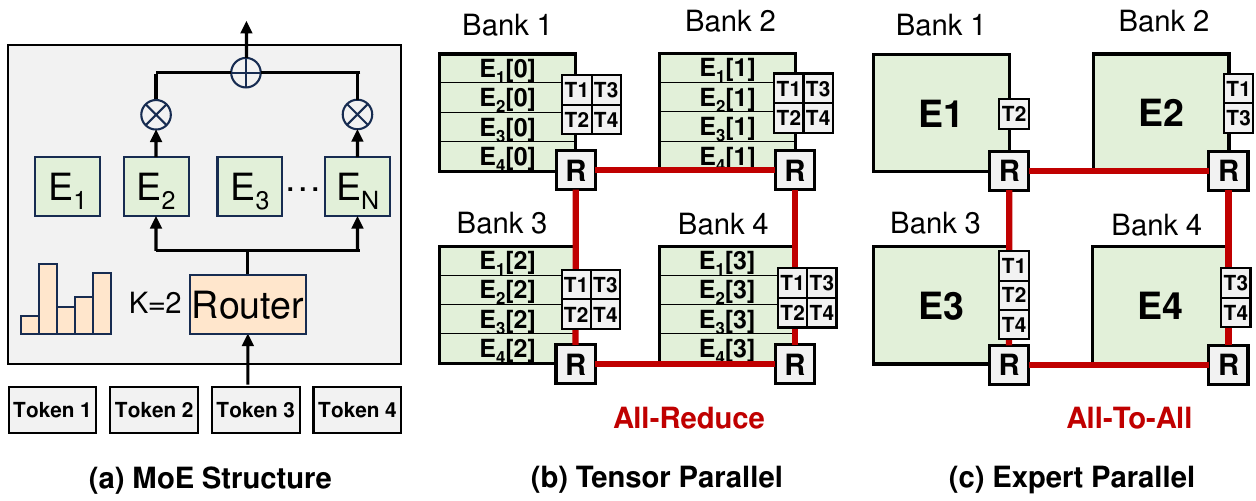}
    \caption{Overview of the MoE architecture and two parallelization strategies.}
    \label{fig:background}
\end{figure}

MoE scales model capacity through sparse execution: a gating network activates only a few experts for each token, reducing per-token computation relative to dense models. Representative MoE models include Mixtral \cite{jiang2024mixtral}, DeepSeek \cite{deepseekv3,deepseekv2}, and Qwen \cite{qwen,yang2024qwen2technicalreport,yang2025qwen3}.

Recent large-scale MoEs typically contain shared and routed experts. Shared experts are activated for every token, yielding deterministic computation that can be assigned statically. Routed experts are selected dynamically according to token features; their irregular activation patterns create the main scheduling and deployment challenges considered in this work.

MoE optimization has been studied across several scenarios. Offloading approaches distribute work across hybrid CPU--GPU platforms \cite{AdapMoE,HybriMoE,zhang2024daop,zhu2025enabling,mcdanel2026moe}. Serving systems improve expert scheduling, load balancing, and placement \cite{eps,moetuner,SCmoe,chen2025eac,li2025speculative}, while training systems optimize combinations of data, tensor, and expert parallelism \cite{lepikhin2020gshard,Switch,FasterMoE,netmoe,lin2025hiermoe}. These studies highlight memory capacity, computation balance, and communication overhead as central deployment constraints.

\subsection{Near-Memory Processing Architectures}

Sparse, bandwidth-intensive MoEs remain limited by the memory wall. Processing-in-Memory colocates computation with data \cite{kim2023samsung}, but its low compute density limits large-model scalability. In contrast, 3D NMP stacks DRAM and logic dies, providing high bandwidth with moderate compute density.

\textbf{Hybrid Bonding DRAM}~\cite{HB1,HB2,similarity} is a representative 3D NMP design whose fine-pitch vertical interconnects provide low-latency, high-bandwidth memory--logic communication. Recent studies further demonstrate 3D NMP for memory-intensive LLM workloads~\cite{huang2025hd,fu2025h,li2025h2,qu20253d,huang2025a3d}.

\subsection{Distributed Inference Strategies}\label{sec:distribute}

Distributed 3D-NMP inference commonly uses Tensor Parallelism (TP) or Expert Parallelism (EP). TP (Fig.~\ref{fig:background}b) splits each expert across processing elements, balancing computation but requiring bandwidth-intensive all-reduce communication that grows with batch size. This communication is difficult to quantize because partial sums use high precision.

EP (Fig.~\ref{fig:background}c) assigns complete experts to PE groups and dispatches tokens to their routed experts. A lightweight metadata all-to-all precedes the dominant hidden-state transfer~\cite{zuo2025serving}, after which outputs are gathered. EP reduces synchronization and enables output quantization, but dynamic routing causes compute imbalance and irregular communication. Fig.~\ref{fig:background} summarizes the trade-off; our design combines TP and EP to balance computation and communication.

%% file: docs/3-motivation.tex
\section{Motivation}

\textcolor{red}{The TP/EP trade-off above is ultimately driven by sparse and dynamic expert routing. We therefore examine three routing-induced characteristics that directly affect MoE deployment on 3D NMP: long-term activation skew, short-term runtime fluctuation, and instantaneous mismatch between model-side routing scores and hardware cost.}

\begin{figure}[!tb]
\centerline{\includegraphics[width=\linewidth]{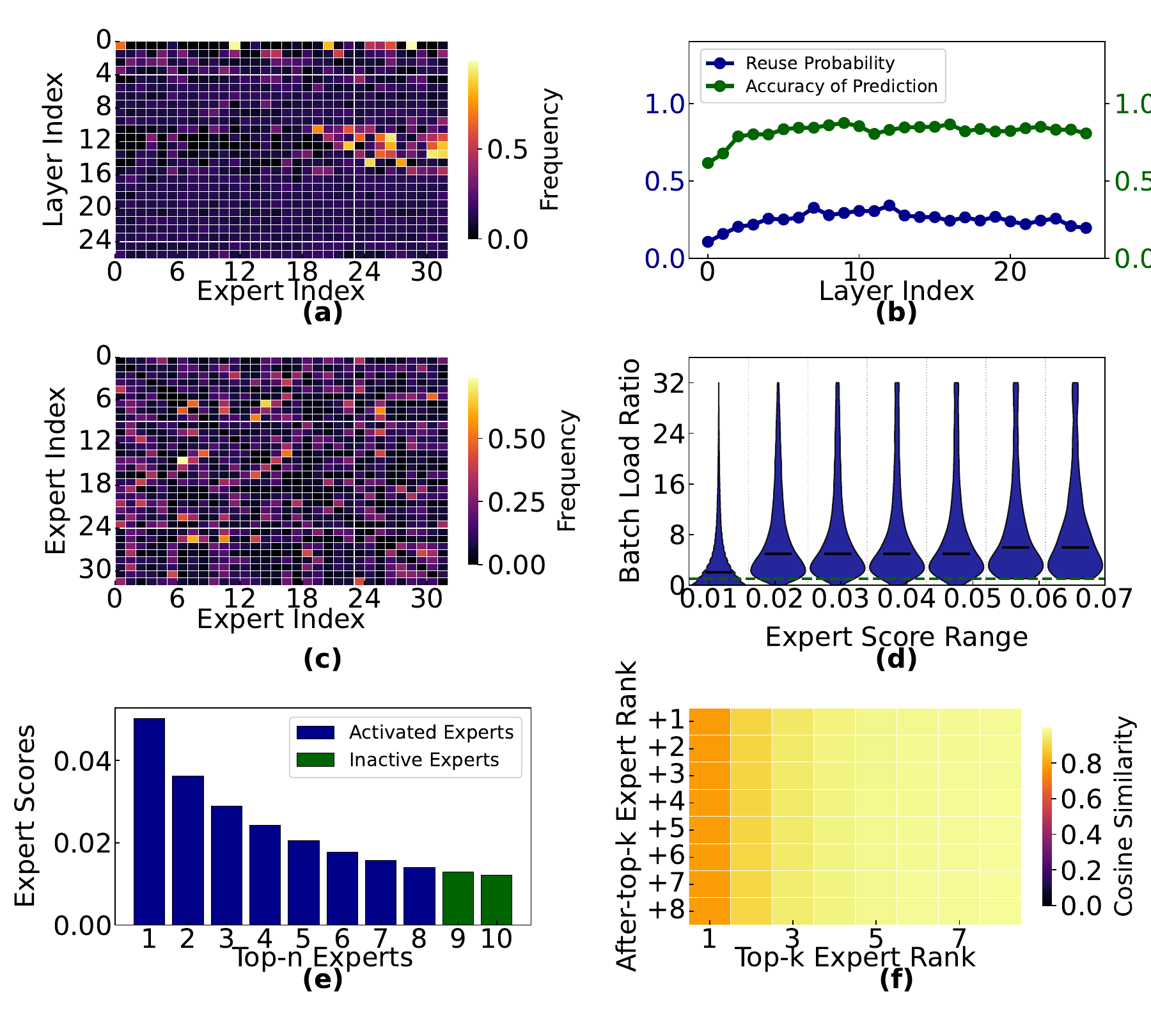}}
\caption{(a) Expert activation frequency, (b) activation overlap between adjacent layers (blue) and next-layer prediction accuracy (green), (c) expert routing affinity, where entry $(i,j)$ denotes the conditional probability that expert $j$ is activated given expert $i$, \textcolor{red}{(d) batch-level load variation among near-boundary expert choices with similar gating scores, where Batch Load Ratio is each expert's number of tokens to process normalized by the batch-average value; 1.0 indicates average load and larger values indicate heavier load, (e) expert score distribution, and (f) output similarity under boundary expert replacement, where entry $(i,j)$ reports the cosine similarity between the original score-weighted top-$k$ MoE output and the output obtained by replacing the $i$-th selected expert with the expert at overall rank $k{+}j$ (the $j$-th after-top-$k$ candidate), while keeping the candidate's original gating weight.}}
\label{fig:motivation}
\end{figure}

\textcolor{red}{\textbf{Long-term routing skew and expert affinity.} Fig.~\ref{fig:motivation}(a) shows that, on Qwen2-57B-A14B-Instruct, a small subset of experts receives many tokens while most experts are rarely activated, complicating expert placement. Fig.~\ref{fig:motivation}(c) further shows strong co-activation affinity. These observations motivate hybrid placement: high-demand experts should be partitioned to avoid hotspots, while low-demand or frequently co-activated experts should be localized to reduce communication.}

\textcolor{red}{\textbf{Short-term runtime fluctuation and predictability.} Fig.~\ref{fig:motivation}(b) shows that expert usage changes noticeably across iterations, which is not captured by a purely static placement. The same figure also shows high next-layer prediction accuracy: due to residual connections, the hidden states of adjacent layers remain similar, so applying the next layer's gating function to the previous layer's hidden states can predict next-layer expert activation accurately. This enables runtime scheduling to proactively allocate resources according to the upcoming routing pattern.}

\textcolor{red}{\textbf{Instantaneous hardware-oblivious routing-cost mismatch.} Standard top-$k$ gating performs token-level expert selection solely according to model-side scores, without considering expert placement, node load, or NoC cost. On Qwen3.5-35B-A3B, Fig.~\ref{fig:motivation}(d) shows that experts with similar boundary scores can have very different batch-level loads, creating a mismatch between model preference and execution cost. To avoid excessive model perturbation, however, hardware-aware routing should avoid changing high-importance expert choices. Fig.~\ref{fig:motivation}(e) shows that only a few activated experts receive dominant gating scores, while lower-ranked selected experts contribute much less to the score-weighted MoE output. Fig.~\ref{fig:motivation}(f) further shows high similarity when lower-ranked selected experts are replaced by near-boundary unselected candidates. These observations motivate hardware-aware gating, which focuses on low-impact boundary choices and limits perturbation to the original routing result.}

%% file: docs/4-method.tex
\section{Performance Modeling}

Before presenting our design, we first developed a performance model to estimate the computation and communication cost of MoE inference on NMP architectures.

% We first build a performance analytical model to estimate the total inference cost, including computation and communication estimation.

\subsection{Computation Overhead Modeling} 

The computation latency $t_{\text{comp}}$ is determined by the most heavily loaded node, and thus reflects the imbalance of expert utilization across the system. For node $c$, its workload depends on the placement matrix $P_{ic}$, where $P_{ic}$ denotes the fraction of expert $i$ assigned to node $c$ (see Table~\ref{parameters} for other notations). The computation time is modeled as
\begin{align}
t_{\text{comp}} = \max_{c} \left\{ \frac{\sum_{i=0}^{E-1} P_{ic} f_i B \cdot 3h \cdot IS}{\text{comp}} \right\}.
\label{comp_load}
\end{align}
Here, $3h \cdot IS$ represents the computation volume per token (including the up, gate, and down projection layer), while $f_i B$ is the number of tokens activating expert $i$. Consequently, $P_{ic} f_i B$ denotes the effective token count processed by node $c$.

% The computation time $t_{\text{comp}}$ is determined by the maximum load across all computing nodes, reflecting expert utilization imbalance. For each node $c$, the load depends on the placement matrix $P_{ic}$, where $P_{ic}$ denotes the proportion that expert $i$ is assigned to node $c$. The explanation of other parameters is listed in the table~\ref{parameters}. The formula is defined as:
% \begin{align}
% t_{\text{comp}} = \max_{c} \left\{ \frac{\sum_{i=0}^{E-1} P_{ic} f_i B \cdot 2h \cdot IS}{\text{comp}} \right\}\label{comp_load}
% \end{align}
% Here, $2h \cdot IS$ is the computation volume per token, $f_iB$ denotes the number of tokens that activate expert $i$, and $P_{ic}f_iB$ is the effective number of tokens that node $c$ needs to process.

% We emphasize that the variables $P_{ic}$ are continuous rather than binary in prior works~\cite{}\hhc{need reference}, allowing our hybrid placement strategy to partially distribute expert computation across multiple nodes (similar to Tensor Parallelism). This design improves deployment flexibility, alleviates hotspots, and enhances compute balance.

% However, this introduces a trade-off with communication overhead. Splitting experts across nodes requires data transfers, which can increase both transfer volume and communication irregularity. Balancing computation load and communication overhead is crucial for optimal MoE performance on 3D PNM architectures.

Note that $P_{ic}$ is modeled as a continuous variable rather than a binary indicator in prior works~\cite{shah2023taccl,moetuner,mei2025helix,zhao2024poster}, allowing experts to be partially distributed across multiple nodes, similar to tensor parallelism. This improves deployment flexibility and alleviates computation hotspots. However, splitting experts also introduces additional communication, creating a trade-off between computation balance and communication cost.

\subsection{Communication Overhead Modeling}\label{comm_model}

% We adopt a discrete-event simulation framework to model irregular all-to-all communications in 2D mesh architectures, extracting the total schedule time as communication overhead.

\textcolor{red}{We model irregular MoE all-to-all communication using a custom discrete-event NoC simulator inspired by the methodology of Ns3~\cite{riley2010ns} and BookSim~\cite{jiang2013detailed}. The simulator generates communication tasks from token--expert mappings, schedules them on directed NoC links, and extracts the total schedule time as communication latency.}

\textcolor{red}{For clarity, schematic figures in this paper mainly use 2D Mesh as the running example, since mesh is the most widely adopted interconnect in prior 3D NMP systems~\cite{huang2025hd,fu2025h,li2025h2,qu20253d,huang2025a3d}. The simulator itself is topology-agnostic: the same abstraction also supports Torus~\cite{dally2004principles} and Fat-tree~\cite{leiserson1985fat} by replacing the path-construction rule and link-bandwidth model.}

\textcolor{red}{To support different physical interconnects, we abstract the NoC as a topology object that provides endpoint lists, candidate paths, path selection, and link bandwidths. For each communication task generated from token--expert routing, the simulator constructs a topology-specific path and schedules the transfer on the directed links along that path. Each directed link maintains an independent occupancy list, and the transfer time is computed using the bandwidth of that specific link.}

\textcolor{red}{Under this abstraction, Mesh uses deterministic dimension-order XY routing with uniform link bandwidth. Torus extends Mesh with wrap-around links and selects the minimum-hop direction in each dimension. Fat-tree uses explicit up-down routes through leaf, aggregation, and core switches, and assigns endpoint--leaf, leaf--aggregation, and aggregation--core bandwidths independently for each directed link. By tracking occupancy and bandwidth for each directed link, the simulator can capture both uniform topologies such as Mesh/Torus and hierarchical topologies such as Fat-tree.}

\textbf{Discrete-Event Simulation for Accurate Latency Estimation.} 
Our simulator generates communication tasks based on token–expert mappings, \textcolor{red}{builds topology-specific paths,} and schedules transmissions using a priority queue while tracking link occupancy to avoid conflicts. Alg.~\ref{comm} summarizes the scheduling procedure.
\begin{algorithm}[H]
\caption{Discrete-Event Simulation for Communication Scheduling}
\label{comm}
\begin{algorithmic}[1] % 启用行号
    \Procedure{SimulateComm}{}
        \State link\_schedule $\gets$ defaultdict(list)
        \State event\_queue $\gets$ []
        \For{each activated expert $e$ of each token $t$}
            \State Find source and destination nodes
            \State \textcolor{red}{path $\gets$ BuildPath(src, dst)}
            \State Add comm\_tasks to event\_queue
        \EndFor
        \State max\_finish\_time $\gets$ 0
        \While{event\_queue}
            \State current\_task $\gets$ heappop(event\_queue)
            \State Find available time on links in path
            \If{link is occupied}
                \State Reschedule task with updated start\_time
            \Else
                \State Remove first link from path
                \State Update current\_time and link\_schedule
                \If{path is not empty}
                    \State Add task back to event\_queue with updated path
                \Else
                    \State max\_finish\_time $\gets$ \texttt{max}(max\_finish\_time, current\_time)
                \EndIf
            \EndIf
        \EndWhile
        \State \Return max\_finish\_time
    \EndProcedure
\end{algorithmic}
\end{algorithm}
\textcolor{red}{In Alg.~\ref{comm}, \texttt{link\_schedule} stores occupied intervals for each directed link. The simulator has three main steps: (1) task generation identifies source nodes holding each token's activated experts and chooses one aggregation destination (lines 4--8); (2) path construction builds the directed-link sequence according to the selected topology (lines 5--6); and (3) task scheduling uses a time-priority heap as the discrete-event queue, always processing the earliest-ready hop first. If the current link is occupied, the event is delayed and reinserted; otherwise, the transfer time is computed from the message size and the current link bandwidth, the occupied interval is appended to \texttt{link\_schedule}, and the task proceeds to the next hop or finishes (lines 10--21).}

\textbf{Linear Approximation for Optimization.} 
% To enable efficient deployment optimization via linear programming (LP), we approximate the communication latency using a node-traffic model: 
To enable efficient deployment optimization, we further approximate the communication latency using a node-traffic model:
\begin{align}
{\color{red}\hat{t}_{\text{comm}} =
\frac{4Bh}{\text{BW}} \max_{c}
\left\{ \sum_{g \in G} f_g
\mathbb{I}\left(\sum_{i\in g}\mathbb{I}(P_{ic}>0)>0\right)
\right\}}
\end{align}
% Here, $\lceil P_{ic} \rceil$ indicates whether expert $i$ is activated on node $c$, and $\prod_{i \in g} \lceil P_{ic} \rceil$ checks if any expert in group $g$ is placed on node $c$. If any expert in group $g$ is placed, the data volume for transfer is $4 f_g B h$, assuming FP32 representation. The total communication volume that node $c$ needs to send is given by $\sum_{g \in G} \left( \prod_{i \in g} \lceil P_{ic} \rceil \right)4 f_g B h$.
\textcolor{red}{The nested indicator $\mathbb{I}\!\left(\sum_{i\in g}\mathbb{I}(P_{ic}>0)>0\right)$ equals 1 if and only if node $c$ hosts at least one expert from co-activated group $g$. Once this condition holds, the tokens associated with group $g$ must communicate with node $c$, producing a communication volume of $4f_gBh$ assuming FP32 representation.}

\textcolor{red}{$\text{BW}$ in Eq.~(2) denotes endpoint-side link bandwidth for the selected topology.}

% We empirically validate the accuracy of the proposed linear approximation model. By comparing the estimated communication latency $\hat{t}_{\text{comm}}$ with the latency values obtained from simulation $t_{\text{comm}}$, we observe a strong linear correlation between the two. The fitting results are shown in Fig.~\ref{communicaton}. 

\begin{figure}[!tb]
\centerline{\includegraphics[width=0.9\linewidth]{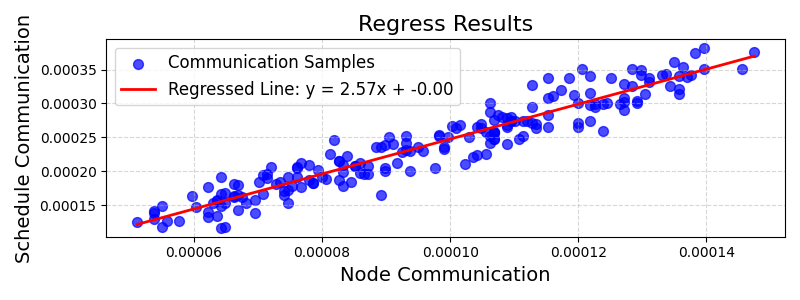}}
\caption{Linear correlation between schedule-based communication latency and node-level communication patterns ($R^2=0.96$).}
\label{communicaton}
\end{figure}

We validate the approximation by comparing the estimated latency $\hat{t}_{\text{comm}}$ with simulation results $t_{\text{comm}}$. As shown in Fig.~\ref{communicaton}, the two exhibit a strong linear correlation:
% As a result, the relationship can be approximated as:
\begin{align}
 t_{\text{comm}} = \gamma \hat{t}_{\text{comm}}
\end{align}
% where $\gamma$ is a scaling coefficient determined through linear regression. Experimental results show that the coefficient of determination ($R^2$) consistently exceeds 0.9 across various scenarios, confirming the model’s reliability in estimating communication overhead for LP optimization.
where $\gamma$ is obtained via linear regression. Across various evaluated scenarios, the coefficient of determination ($R^2$) exceeds 0.9, indicating that the model provides reliable estimates.

% In structured communication patterns like ring-based all-reduce \cite{ring}, commonly used in TP training and inference, the communication algorithm is regular and deterministic. In such cases, total communication time nearly equals per-node communication time due to the algorithm's balanced nature.

For structured communication patterns such as ring all-reduce \cite{ring}, the communication schedule is deterministic and balanced, making the total latency close to the per-node communication time.
% In this case, total communication time nearly equals per-node communication time due to the algorithm's balanced nature.
\begin{align}
 t_{\text{comm}} \approx  \hat{t}_{\text{comm}} \approx \frac{4Bh}{\text{BW}}
\end{align}
% This implies that $\gamma = 1$ for ring all-reduce, demonstrating that our linear communication model is not only accurate for irregular all-to-all traffic, but also directly applicable to structured communication paradigms such as tensor parallel all-reduce.
This implies that $\gamma=1$, showing that the model is also applicable to structured communication such as TP all-reduce.

\textcolor{red}{Accordingly, $\hat{t}_{\text{comm}}$ uses this endpoint-side bandwidth for all topologies, while the discrete-event simulator uses the bandwidth of each traversed directed link directly. Differences in routing paths and link contention under each topology are reflected by refitting $\gamma$ from simulated latency samples when the topology or bandwidth setting changes.}

% We validate our analytical model by comparing its results with the widely used distributed deep learning simulator ASTRA-sim \cite{9238637} for the ring all-reduce operation. As shown in the table \ref{tab:astra_sim}, the latency predictions from the analytical model closely match the simulation results, demonstrating strong alignment between the two. This confirms that our analytical model provides a reliable estimate of latency and can be effectively used for performance prediction and optimization in similar distributed systems.

Finally, we compare the performance model with results from the widely used distributed deep learning simulator ASTRA-sim \cite{9238637}. As shown in Table~\ref{tab:astra_sim}, the predicted latency aligns with the simulation results within \textcolor{red}{3.2\%} relative error across the evaluated ring all-reduce settings, confirming the accuracy of the proposed model.

\begin{table}[!tb]
\centering
\caption{Performance-model validation against ASTRA-sim.}

\begin{center}
\resizebox{0.45\textwidth}{!}{ % 调整宽度为 80% 文本宽度，高度自动缩放

    \begin{tabular}{c|c|c|c|c}
    \hline\hline
    \textbf{Latency} & \textbf{Bandwidth} & \textbf{Predicted Latency} & \textbf{ASTRA-sim Latency} & \textcolor{red}{\textbf{Error}} \\
    \hline
    \textcolor{red}{0.1 us} & \textcolor{red}{25 Gb/s} & \textcolor{red}{673 us} & \textcolor{red}{668 us} & \textcolor{red}{0.7\%} \\
    \hline
    \textcolor{red}{5 us} & \textcolor{red}{25 Gb/s} & \textcolor{red}{851 us} & \textcolor{red}{879 us} & \textcolor{red}{3.2\%} \\
    \hline
    \textcolor{red}{0} & \textcolor{red}{20 Gb/s} & \textcolor{red}{671 us} & \textcolor{red}{692 us} & \textcolor{red}{3.0\%} \\
    \hline
    \textcolor{red}{0.1 us} & \textcolor{red}{20 Gb/s} & \textcolor{red}{705.6 us} & \textcolor{red}{691.5 us} & \textcolor{red}{2.0\%} \\
    \hline
    \textcolor{red}{0.1 us} & \textcolor{red}{50 Gb/s} & \textcolor{red}{285.6 us} & \textcolor{red}{281.3 us} & \textcolor{red}{1.5\%} \\
    \hline
    \textcolor{red}{0.1 us} & \textcolor{red}{10 Gb/s} & \textcolor{red}{1405.6 us} & \textcolor{red}{1375.1 us} & \textcolor{red}{2.2\%} \\
    \hline\hline
    \end{tabular}}
    \label{tab:astra_sim}
\end{center}
\end{table}

\begin{table}[!tb]
\centering
\caption{Main notations.}
\begin{center}
\resizebox{0.45\textwidth}{!}{ % 调整宽度为 80% 文本宽度，高度自动缩放
\begin{tabular}{c|c}

\hline\hline
\textbf{Notation} & \textbf{Definition} \\
\hline
$t_{\text{comp}}$ & Computation latency \\
\hline
$t_{\text{comm}}$ & Communication latency \\
\hline
$\hat{t}_{\text{comm}}$ & Approximated communication latency \\
\hline
$c$ & Node index \\
\hline
$E$ & Total number of experts \\
\hline
$e$ & Number of activated experts per token \\
\hline
$P_{ic}$ & Fraction of expert $i$ assigned to node $c$ \\
\hline
$f_i$ & Activation frequency of expert $i$ \\
\hline
$B$ & Batch size \\
\hline
$h$ & Hidden dimension \\
\hline
$IS$ & MoE intermediate size \\
\hline
$D$ & Number of nodes \\
\hline
$\text{comp}$ & Per-node compute throughput \\
\hline
\textcolor{red}{$\text{BW}$} & \textcolor{red}{Endpoint-side link bandwidth} \\
\hline
\textcolor{red}{$M_i$} & \textcolor{red}{Expert-$i$ weight size} \\
\hline
\textcolor{red}{$\rho_{\text{mem}}$} & \textcolor{red}{DRAM capacity cap factor} \\
\hline
$G$ & Set of co-activated expert groups \\
\hline
$g$ & Expert-group index \\
\hline
$f_g$ & Co-activation frequency of expert group $g$ \\
\hline
\textcolor{red}{$\mathbb{I}(\cdot)$} & \textcolor{red}{Indicator function} \\
\hline
\textcolor{red}{$Y_{gc}$} & \textcolor{red}{Group-node indicator in LP} \\
\hline
$\mathbf{G}(\mathbf{x})$ & Adjusted gating score vector \\
\hline
$W_g$ & Gating weight matrix \\
\hline
$b_g$ & Gating bias vector \\
\hline
$r_{\text{comp}}$ & Computation penalty coefficient \\
\hline
$r_{\text{comm}}$ & Communication penalty coefficient \\
\hline
$\mathbf{T}_{\text{comp}}$ & Computation penalty vector \\
\hline
$\mathbf{T}_{\text{comm}}$ & Communication penalty vector \\
% \hline
% $T_i$ & Penalty of expert $i$ \\
% \hline
% $\mathbf{P}_i$ & Placement vector of expert $i$ across nodes \\
% \hline
% $\mathbf{1}_E$ & All-one vector of length $E$ \\
% \hline
% $I_E$ & Identity matrix of size $E$ \\
\hline\hline
\end{tabular}}
\label{parameters}
\end{center}
\end{table}

\section{\method~ Design}\label{sec:method}

\subsection{Overview}

% Fig. \ref{overview} provides an overview of the \method~ framework, which consists of an offline mapping phase and an online inference phase. The offline phase involves automated hybrid expert mapping on 3D NMP with Node Balance (Sec. \ref{node balance}) and Link Balance (Sec. \ref{link balance}) optimization techniques. During online inference, dynamic scheduling is employed to predict computation load, prioritize experts (Sec. \ref{prediction}), and pre-broadcast the expert with highest load (Sec. \ref{broadcast}), ensuring efficient resource utilization without inducing additional communication overhead (Sec. \ref{communication-eff}). In addition, Hardware-aware Gating (Sec.~\ref{adaptive-gating}) is applied to further guide expert routing decisions toward more balanced and communication-efficient execution.

\begin{figure}[!tb]
\centerline{\includegraphics[width=\linewidth]{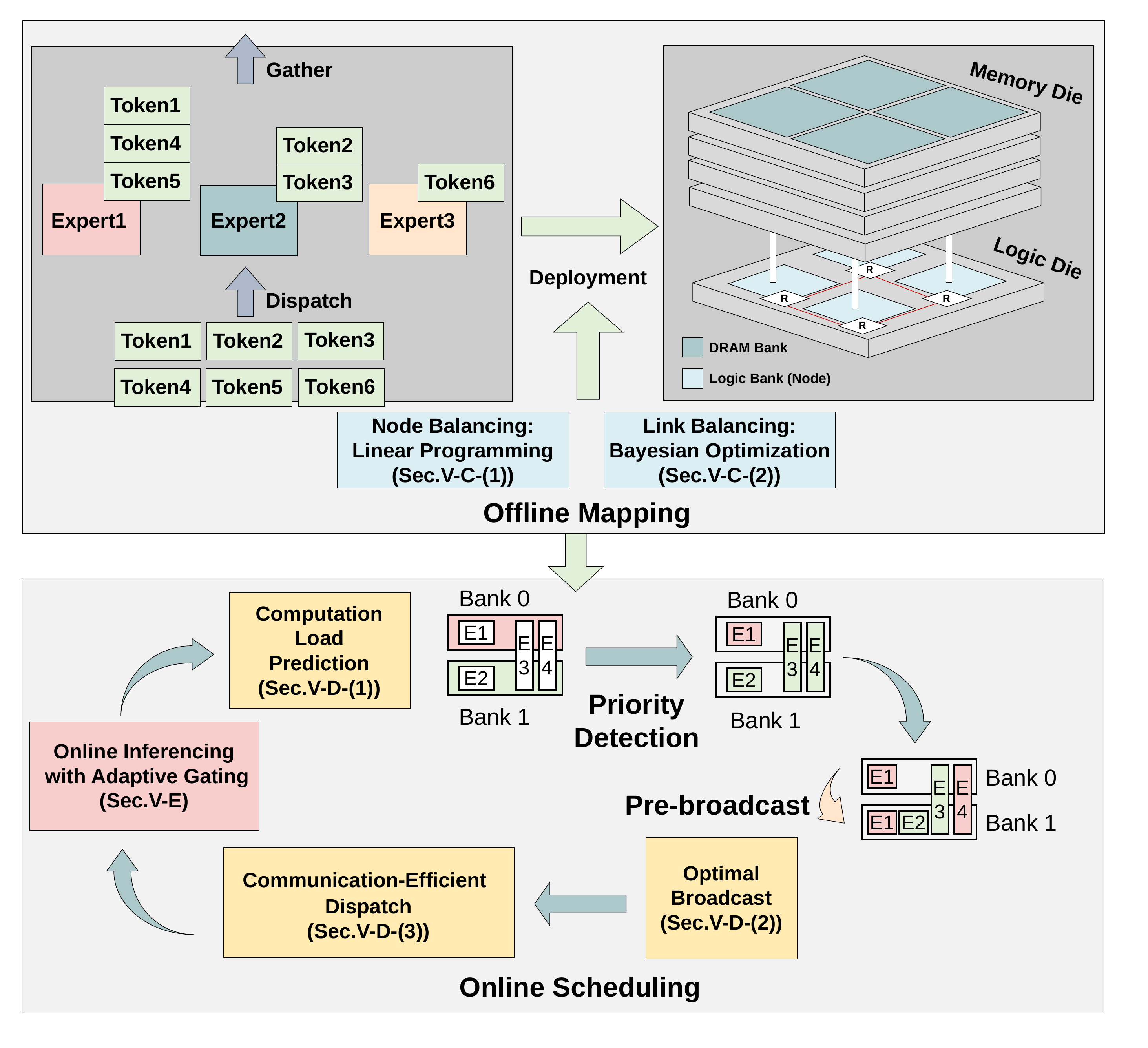}}
\caption{Overview of \method}
\label{overview}
 \end{figure}

Fig.~\ref{overview} illustrates \method, which includes offline placement and online inference. Offline placement determines hybrid expert placement through Node Balance (Sec.~\ref{node balance}) and Link Balance (Sec.~\ref{link balance}).
\textcolor{red}{The pipeline follows the bottleneck-latency objective in Sec.~IV and applies it at three decision granularities. At placement granularity, Node Balance reduces the maximum node compute load and the node-level communication proxy. At physical-mapping granularity, Link Balance maps the logical placement to physical endpoints to reduce link congestion.}
\textcolor{red}{At runtime granularity, the online stage applies latency-model-derived local decisions from two complementary sides. Dynamic scheduling operates on the hardware supply side: it changes how already-routed tokens use available expert replicas by adjusting pre-broadcast and dispatch decisions, without modifying routing scores. Hardware-aware gating operates on the software demand side: it slightly adjusts low-impact expert choices so that token routing itself requests less congested computation and communication resources. Together, they optimize online execution from two different angles under the same latency model, with their effects validated in Sec.~VI.}

\subsection{\textcolor{red}{Automated Hybrid Parallelism via Node-Link Balance}}\label{sec:hybrid-parallelism}

% We propose a two-stage \textbf{Node-Link Balance Co-optimization} strategy for efficiently deploying MoE models on 3D NMP architectures. The placement problem is divided into a logical optimization stage that balances computational load and reduces communication volume, and a physical mapping stage to minimize link-level congestion. This separation of logical workload and physical interconnects simplifies the placement problem, as detailed in the following stages.
\textcolor{red}{Automated hybrid parallelism jointly decides how experts are split across $D$ logical clusters and how these clusters map to physical NoC endpoints, using the model configuration, workload profile, and hardware setting as inputs.} We adopt a two-stage \textbf{Node-Link Balance Co-optimization} strategy for MoE deployment on 3D NMP architectures. The first stage optimizes logical expert placement to balance computation and reduce communication volume, while the second stage maps logical clusters onto physical nodes to alleviate link congestion. This decomposition simplifies the placement and enables efficient co-optimization of computation and communication.

\textcolor{red}{Alg.~\ref{alg:hybrid-parallelism} summarizes the overall search procedure.}
\begin{algorithm}[H]
\caption{\textcolor{red}{Automated Hybrid Parallelism Search}}
\label{alg:hybrid-parallelism}
\small
\begingroup
\color{red}
\begin{algorithmic}[1]
    \Require modelConfig, workloadProfile, hardwareConfig
    \Ensure hybrid placement $\{P^{(\ell)}\}$, mapping $\{M^{(\ell)}\}$
    \State $(G,\{f_i\},\{f_g\}) \gets$ \Call{ExtractStats}{workloadProfile}
    \State $\gamma \gets$ \Call{CalibrateProxy}{hardwareConfig}
    \For{each MoE layer $\ell$}
        \State $P^{(\ell)} \gets$ \Call{NodeBalanceLP}{$\ell$}
        \State $M^{(\ell)} \gets$ \Call{LinkBalanceBO}{$P^{(\ell)}$}
    \EndFor
\end{algorithmic}
\endgroup
\end{algorithm}
\subsubsection{Node Balance Optimization via Linear Programming}\label{node balance}
% In the first stage of our co-optimization framework, we focus on balancing computational and communication workloads across logical compute clusters, abstracting away their physical topology. Given the large number of variables and the combinatorial nature of the expert placement problem, manual tuning becomes infeasible. Therefore, we adopt a linear programming (LP) formulation to enable automated and scalable optimization across diverse hardware configurations and inference scenarios. The LP model simultaneously considers computation bottlenecks and approximated communication costs using the linear estimator $\hat{t}_{\text{comm}}$ derived earlier. 
In the first stage, we optimize expert placement across logical compute clusters, abstracting away physical routing and using the configured bandwidth in the communication proxy. Since the placement space is highly combinatorial, manual tuning becomes infeasible. Therefore, we formulate the problem as a linear program (LP). The LP jointly models computation bottlenecks and approximated communication cost using the estimator $\hat{t}_{\text{comm}}$ derived earlier.

% The notations of the LP formulation are defined in Table~\ref{parameters}. Among them, the continuous variables $P_{ic} \in [0, 1]$ represent the proportion of expert $i$’s workload assigned to cluster $c$. The binary decision variables $Z_{ic} \in \{0, 1\}$ indicates whether expert $i$ is active on cluster $c$.
The notations are defined in Table~\ref{parameters}. The continuous variable $P_{ic}\in[0,1]$ denotes the fraction of expert $i$ assigned to cluster $c$, the binary variable $Z_{ic}\in\{0,1\}$ indicates whether expert $i$ is placed on cluster $c$, and \textcolor{red}{$Y_{gc}=\mathbb{I}\left(\sum_{i\in g}Z_{ic}>0\right)$ indicates whether cluster $c$ stores at least one expert from co-activated group $g$}. 
\textcolor{red}{Because $P_{ic}$ is continuous, an expert can either stay on one cluster (EP-style) or be partitioned across clusters (TP-style). Thus, \method~ automatically decides the per-expert parallelism mode through the optimized $P_{ic}$ and $Z_{ic}$, instead of manually labeling experts as TP or EP.}
Then we define some constraints to guarantee the legal mapping and efficiently search for the optimal allocation strategy.
\begin{align}
    & Z_{ic} \ge P_{ic},  \quad \forall (i,c) & \label{eq:const1}\\
    & {\color{red}Y_{gc} \ge Z_{ic}, \quad \forall g, i\in g, c} & \label{eq:constY}\\
    & \sum_{c} P_{ic} = 1,\quad  \forall i  & \label{eq:const4}\\
    & {\color{red}t_{\text{comm}} \ge \gamma\frac{4Bh}{\text{BW}} \sum_{g\in G} f_gY_{gc},\quad  \forall c}  & \label{eq:const3}\\
    & t_{\text{comp}} \ge \frac{\sum_{i=0}^{E-1}P_{ic}f_iB \cdot 3h \cdot IS}{\text{comp}},\quad  \forall c  & \label{eq:const2}\\
    & {\color{red}\sum_{i=0}^{E-1} P_{ic}M_i \leq \rho_{\text{mem}}\frac{\sum_{i=0}^{E-1}M_i}{D},\quad \forall c} & \label{eq:memcap}\\
    & 0 < \sum_{i=0}^{E-1} P_{ic}f_i   \leq \left( \frac{1}{R_{CC}} + 1 \right) \frac{ e}{D},\quad  \forall c & \label{eq:const5} \\
    & R_{CC} = \frac{t_{\text{comp}}}{t_{\text{TP,comm}}} = \frac{\text{BW} \cdot IS \cdot e}{2D \cdot \text{comp}} & \label{eq:const6}
\end{align}

% Constraints \ref{eq:const1} and \ref{eq:const4} handle expert placement and workload assignment across nodes. Constraints \ref{eq:const2} and \ref{eq:const3} ensure balanced computation and communication times, minimizing bottlenecks.

% Constraints \ref{eq:const5} and \ref{eq:const6} restrict node workload to avoid imbalance, with an upper bound set by the theoretical compute + communication time of TP inference. These constraints prune suboptimal placements early, reducing search space complexity and improving solver convergence without sacrificing optimality.
\textcolor{red}{Constraints~\ref{eq:const1}--\ref{eq:const4} define legal expert assignment and group-node indicators. Constraints~\ref{eq:const3} and \ref{eq:const2} upper-bound the communication and computation bottlenecks, respectively.}
In particular, Constraint~\ref{eq:const3} models communication at the granularity of co-activated expert groups, where each $g\in G$ denotes a group of experts activated together, and $f_g$ represents its co-activation frequency. \textcolor{red}{It is the LP linearization of Eq.~(2): the nonlinear nested indicator in Eq.~(2) is replaced by the auxiliary binary variable $Y_{gc}$ through Constraint~\ref{eq:constY}, while the factor $\gamma$ applies the calibrated linear approximation from Eq.~(3). By co-locating frequently co-activated experts, the optimization reduces the number of nodes involved in each active group and thus lowers the communication term $\sum_{g\in G} f_gY_{gc}$.}

\textcolor{red}{Expert placement is also limited by each 3D NMP node's local memory capacity. Constraint~\ref{eq:memcap} captures this requirement by bounding the expert-weight storage assigned to each node. Here $M_i$ is the weight size of expert $i$, and $\rho_{\text{mem}}$ controls how much the maximum per-node memory load may exceed the uniform average. A smaller $\rho_{\text{mem}}$ enforces more balanced storage, while a larger value gives the latency-oriented objective more placement freedom. This prevents the LP from concentrating too many expert weights on a few nodes, improving memory-capacity utilization while preserving the latency objective.}
Constraints~\ref{eq:const5} and \ref{eq:const6} restrict node workload using the theoretical TP compute--communication ratio, which prunes suboptimal placements and improves solver convergence.

Finally, we minimize the node-level inference overhead as
\begin{align}
&\min t_{\text{node\_overhead}}\\
&t_{\text{node\_overhead}}=t_{\text{comp}}+2t_{\text{comm}}=t_{\text{comp}}+2\gamma \hat{t}_{\text{comm}}
\end{align}
% Here, $\gamma$ is the scaling coefficient empirically derived from simulation (Sec. \ref{comm_model}), and the factor 2 accounts for the cost of all-to-all dispatch and all-to-all gather, which constitute a pair of symmetric communication operations.

% This LP formulation enables a globally coordinated placement strategy that balances computation and communication, providing a foundation for the second stage of physical mapping.
Here, $\gamma$ is the scaling factor obtained from the communication model (Sec.~\ref{comm_model}), and the factor 2 accounts for the symmetric all-to-all dispatch and gather. This LP yields a globally coordinated logical placement that balances computation and communication, providing the basis for the subsequent physical mapping stage.
\subsubsection{Link Balancing via Bayesian Optimization}\label{link balance}

% In this stage, the logical clusters are mapped to physical nodes on the 2D mesh network. The objective here is to minimize link congestion and improve communication tail latency. We adopt Bayesian Optimization to search for low-congestion mapping strategies, as it is well-suited for problems with expensive evaluations and relatively smooth objective functions—e.g., swapping nearby clusters causes only minor changes in communication cost. This enables efficient exploration of the mapping space with minimal simulation overhead.
In the second stage, the optimized logical clusters are mapped onto physical nodes in the 2D mesh. The goal is to reduce link congestion and communication tail latency. We use Bayesian Optimization to search for low-congestion mappings, as the objective is expensive to evaluate but changes smoothly under local mapping variations (e.g., swapping nearby clusters causes only minor changes in communication cost).

\textcolor{red}{The same Link Balance procedure extends to Torus and Fat-tree with the topology-specific routing and bandwidth defined in Sec.~\ref{comm_model}. For Mesh and Torus, the goal is to reduce communication congestion. For Fat-tree, it aims to reduce cross-leaf, cross-pod, and core-level communication.}

\begin{figure}[!tb]
\centerline{\includegraphics[width=\linewidth]{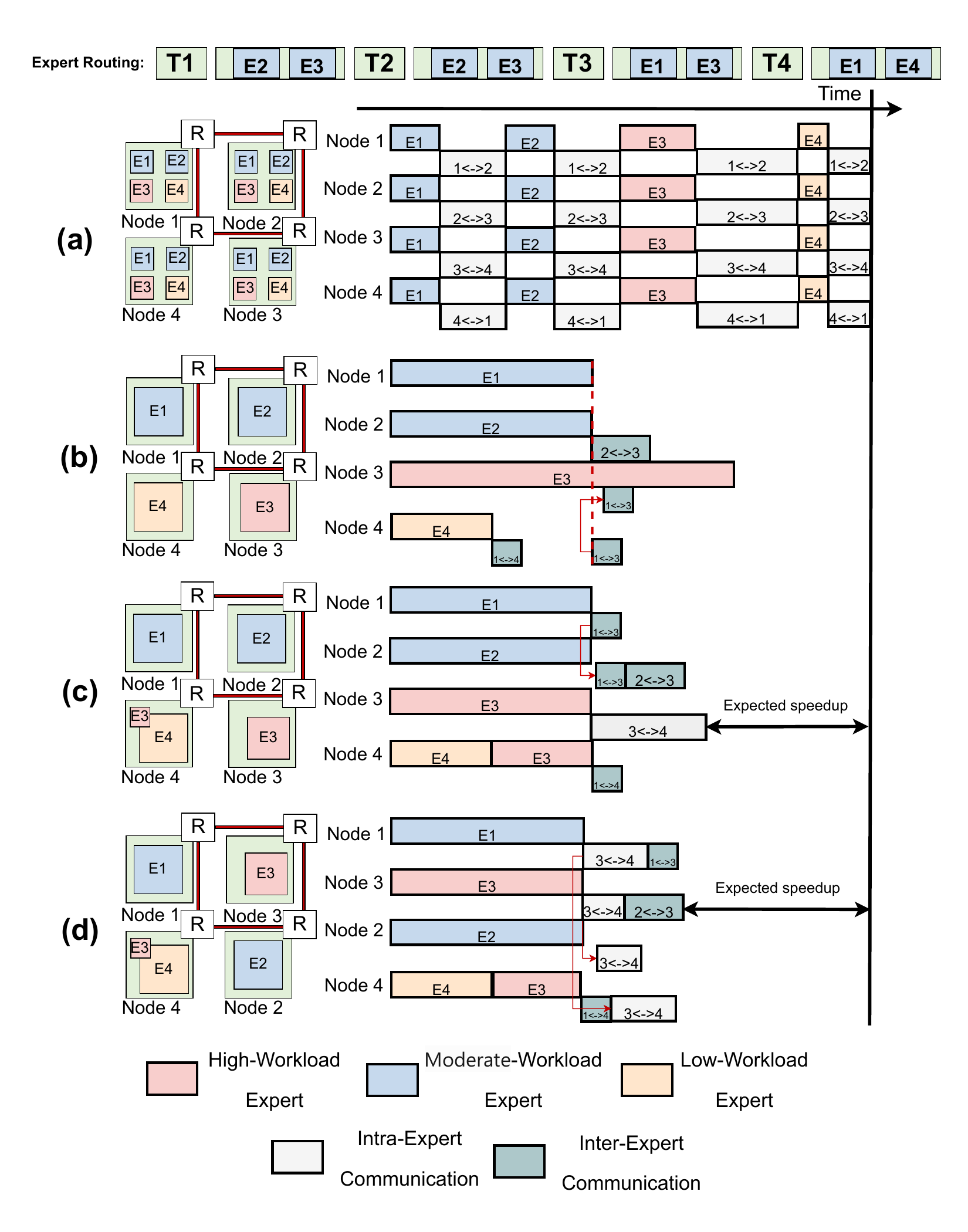}}
\caption{(a) TP: balanced computation but communication-intensive; (b) EP: communication-efficient but computation-imbalanced, (c) Hybrid parallel with node balance: balanced computation with irregular traffic; (d) Hybrid parallel with node-link balance: balanced computation with regular and less congested traffic.}
\label{node-link}
\end{figure}

% Fig.~\ref{node-link} illustrates example placements under four parallelism strategies, highlighting the trade-offs between computation balance and communication overhead in MoE inference. MoE communication can be categorized into Intra-Expert Communication, where a single expert is split across nodes and requires result aggregation (as in Tensor Parallelism), and Inter-Expert Communication, where multiple experts activated by the same token need to exchange and aggregate results. In (a), TP achieves balanced computation by splitting all experts across nodes, but incurs heavy Intra-Expert communication as each token triggers all-reduce operations. In (b), EP avoids Intra-Expert communication, with only sparse Inter-Expert communication—e.g., between E1 (Expert 1) and E3 via Node 4. However, tokens T1, T2, and T3 all activate E3, leading to overload on Node 3 and poor resource utilization. In (c), Node Balance optimization redistributes part of E3 to Node 4 to balance computation. Yet, this split introduces Intra-Expert communication between Nodes 3 and 4, causing link congestion along the 3→4 path while leaving other links underutilized. In (d), Node-Link Balance adjusts the physical mapping (e.g., swapping Node 2 and Node 3), allowing the synchronization between Nodes 3 and 4 to be routed via 3→1→4 and 3→2→4, which alleviates link congestion and achieves both balanced computation and regular, uncongested communication.
Fig.~\ref{node-link} illustrates the placement trade-off. TP balances computation but introduces heavy intra-expert synchronization; EP avoids such synchronization but can overload hot experts. Node Balance splits overloaded experts to reduce compute hotspots, but the induced synchronization may congest specific paths. Node-Link Balance further remaps logical clusters onto physical nodes, spreading synchronization traffic over less congested links and achieving both computation and communication balance.
In the example, tokens T1--T3 all activate E3, so EP overloads Node 3. Node Balance splits E3 across Nodes 3 and 4 to remove the compute hotspot, while Node-Link Balance further remaps physical nodes so the induced synchronization avoids a single congested link.

\subsection{Dynamic Scheduling}\label{sec:dynamic}
% We propose a runtime-adjustable deployment strategy for dynamic expert routing in MoE inference, consisting of three key components: congestion-aware expert prediction, cost-optimal broadcasting, and communication-efficient token routing.
We design a runtime scheduling strategy for dynamic expert routing in MoE inference, including three components: expert priority prediction, cost-aware pre-broadcasting, and communication-efficient token dispatch.

\subsubsection{Priority Detection and Computation Prediction}\label{prediction}

% We leverage temporal locality in expert activations to predict computation hotspots in the next layer. For each expert $i$ on node $c$, we define a priority score that estimates its future compute cost:
Motivated by the similarity of adjacent-layer hidden states induced by residual connections, we predict next-layer computation hotspots by applying the next layer's gate to the previous layer's hidden states. For expert $i$ on node $c$, its priority score is defined as
\begin{align}
&prio_{ic}=\frac{3P_{ic}\hat{f}_i\cdot IS}{comp}
\end{align}
\textcolor{red}{This score ranks experts by their predicted marginal contribution to the maximum node compute load in Eq.~\eqref{comp_load}, i.e., the expected latency benefit of pre-broadcasting them under the current hotspot.}
% Here, $\hat{f}_i$ is the predicted activation frequency of expert $i$. The expert with the highest priority on the most congested node is selected for pre-broadcast, repeating this process for a limited number of iterations based on the previous layer’s inference latency.

% Here, $\hat{f}_i$ denotes the predicted activation frequency of expert $i$. The expert with the highest priority on the most congested node is selected for pre-broadcast, and this process is repeated for a limited number of iterations determined by the previous layer's latency.
Here, $\hat{f}_i$ denotes the predicted activation frequency of expert $i$. The highest-priority expert on the most congested node is pre-broadcast repeatedly within the runtime budget of the previous layer. The required metadata, such as priority scores and dispatch tables, are batch-local and can be piggybacked on the lightweight metadata exchange before token dispatch (Sec.~\ref{sec:distribute}), so the overhead is negligible.

\subsubsection{Optimal Broadcast Chunk Size}\label{broadcast}

% Broadcasting an expert involves splitting it into chunks of size $c$, with the following trade-offs:

% \begin{itemize}
%     \item Larger chunks reduce hops, lowering latency.
%     \item Smaller chunks reduce per-hop traffic but increase transmission delays.
% \end{itemize}

% The traditional $\alpha$–$\beta$ communication model can clearly describe this kind of trade-off:

Pre-broadcasting an expert splits its parameters into chunks of size $c$. Larger chunks reduce the number of hops and thus latency, while smaller chunks reduce per-hop traffic but increase transmission delay. This trade-off can be modeled by the traditional $\alpha$--$\beta$ communication model:
\begin{align}
&\text{latency}=\alpha (2\sqrt{D}+\frac{h\cdot IS}{c})\\
&\text{bandwidth}=\beta (h\cdot IS+2c\sqrt{D})\\
&t_{\text{pre\_b}}=\text{latency}+\text{bandwidth}
\end{align}
% Here, $k$ is the number of pre-broadcast iterations allowed within the runtime window. The model yields a lower bound:
Let $k$ denote the number of pre-broadcast iterations allowed within the runtime window. Then the lower bound of $t_{\text{pre\_b}}$ is
\begin{align}
    t_{\text{pre\_b}}\ge h\cdot IS\cdot\beta k+2\alpha \sqrt{D}+2\sqrt{2\sqrt{D}\beta k\alpha h\cdot IS}
\end{align}
% This bound is tight when the chunk size $c$ is selected optimally as:
which is tight when the chunk size is chosen as
\begin{align}
    c=\sqrt{\frac{\alpha h\cdot IS}{2\beta k\sqrt{D}}}
\end{align}
% This provides a solution for the most efficient pre-broadcast under a given runtime window constraint.
This gives the optimal chunk size for efficient pre-broadcast under the runtime constraint.
\subsubsection{Communication-Efficient Dispatch}\label{communication-eff}
% After expert broadcasting, each token can be routed to any node holding a copy of its activated experts. To avoid incurring additional communication overhead, we restrict the routing candidates to nodes where the routed experts are already present. Among these candidates, the token is dispatched to the node with the lowest current compute load, minimizing workload imbalance without introducing extra data movement.
After broadcasting, a token can be routed to any node that already holds its activated experts. To avoid extra communication, we restrict the candidate nodes to those containing the routed experts and select the one with the lowest current compute load. This reduces workload imbalance without introducing additional data movement.

\begin{figure}[!tb]
\centerline{\includegraphics[width=\linewidth]{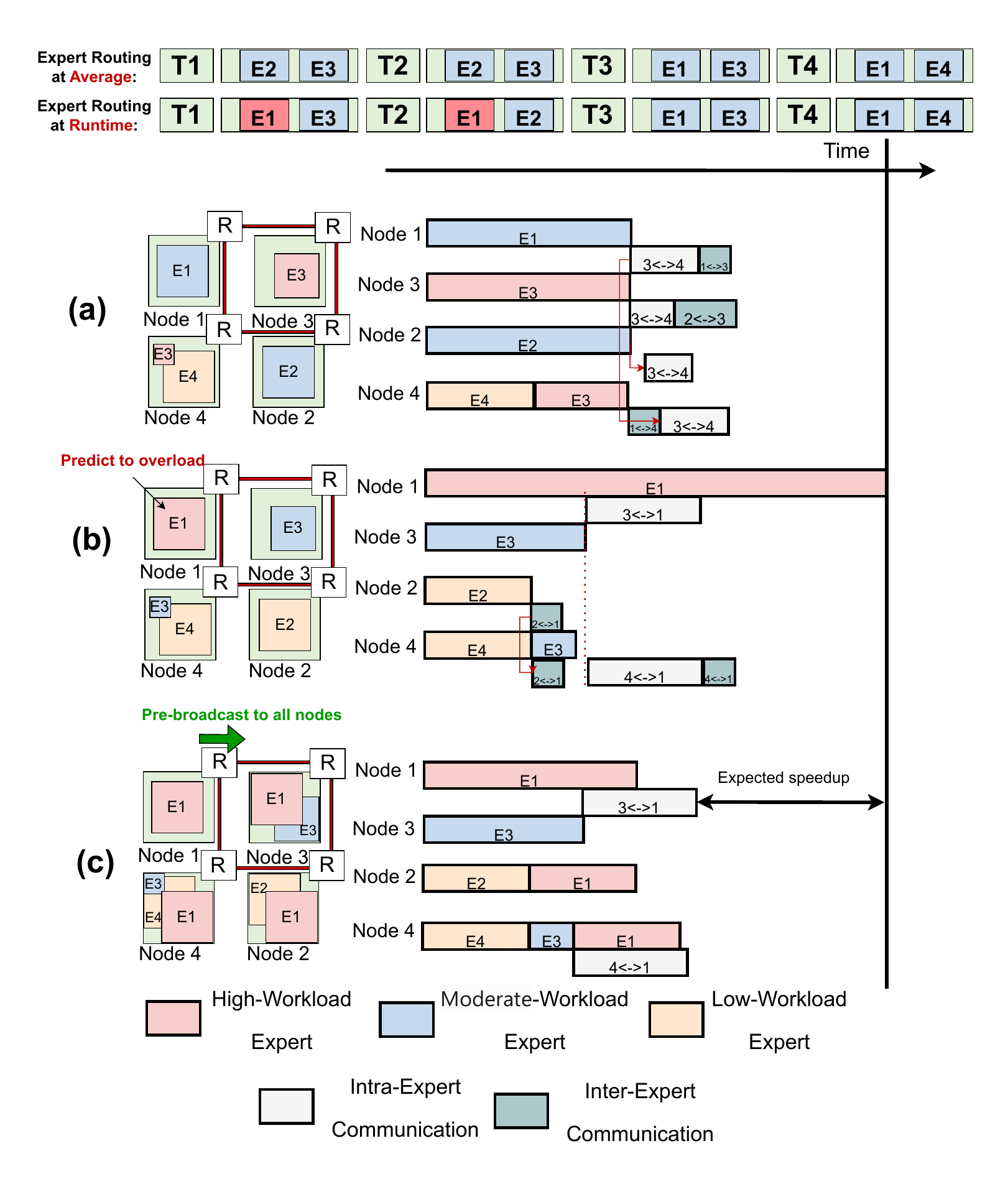}}
\caption{(a) Static placement, (b) hotspot identification, (c) pre-broadcast of the hottest expert and token dispatch without extra communication.}
\label{fig:dynamic}
 \end{figure}

% An example of dynamic expert scheduling is illustrated in Fig.~\ref{fig:dynamic}. Subfigure (a) shows the static deployment from Fig.~\ref{node-link}, which performs efficiently under averaged expert activation patterns. However, during real inference, as shown in (b), expert activation becomes highly skewed—Expert 1 (E1) turns into a bottleneck due to concentrated token routing. Our priority detection mechanism can anticipate such overload at runtime, identifying E1 as a high-demand expert in advance. As shown in (c), E1 is then \textbf{pre-broadcast} to all nodes before execution. Tokens such as T2 and T4 are routed to Node 2 and Node 4 respectively, both holding E1, thus avoiding additional Inter-Expert communication. This not only balances the computation load but also improves communication efficiency.
Fig.~\ref{fig:dynamic} illustrates dynamic scheduling: when E1 becomes a runtime hotspot, it is pre-broadcast to all nodes, allowing tokens such as T2 and T4 to execute on nodes already holding E1 and thus improving load balance without extra inter-expert communication.
Specifically, Fig.~\ref{fig:dynamic}(b) identifies E1 as the predicted hotspot, and Fig.~\ref{fig:dynamic}(c) dispatches the routed tokens to existing E1 copies instead of moving expert outputs across nodes.
\textcolor{red}{This online scheduling rule is greedy: finding the globally optimal broadcast and dispatch plan would require solving a discrete scheduling problem under the current link and compute queues during every runtime window. Such optimization would exceed the available online budget, so we rank candidate broadcasts and dispatches using the priority score derived from the validated latency model in Sec.~\ref{comm_model}.}

\subsection{Hardware-aware Gating}\label{adaptive-gating}

While dynamic scheduling reduces latency by pre-broadcasting hot experts, runtime bandwidth and pre-broadcast budget may remain insufficient when multiple hotspots coexist, especially at small batch sizes.
\textcolor{red}{Hardware-aware gating therefore complements dynamic scheduling on the software demand side. Dynamic scheduling increases the usable hardware supply for already-selected experts, whereas hardware-aware gating incorporates marginal computation and communication penalties into expert routing to reduce the resources requested by token routing. These penalties are derived from the latency model in Sec.~IV.}
\subsubsection{Objective}
\textcolor{red}{Hardware-aware gating implements routing with latency regularization: the softmax term preserves model utility, while $\mathbf{T}_{\text{comp}}$ and $\mathbf{T}_{\text{comm}}$ represent the marginal system cost of selecting each expert under the current placement and scheduling state.}
Specifically, we redefine the gating function as follows:
\begin{align}
&\mathbf{G}(\mathbf{x}) =\text{Softmax}(W_g \mathbf{x} + b_g)+r_{\text{comp}}\mathbf{T}_{\text{comp}}+r_{\text{comm}}\mathbf{T}_{\text{comm}}
\end{align}
where $\mathbf{T}_{\text{comp}}$ and $\mathbf{T}_{\text{comm}}$ denote the computational and communication overhead, respectively, associated with selecting each expert. The hyperparameters $r_{\text{comp}}$ and $r_{\text{comm}}$ are typically negative-valued and act as penalty terms, controlling the trade-off between system efficiency and model fidelity.

The runtime overhead of hardware-aware gating is small, since the penalty terms are computed through lightweight matrix operations and reductions over routing metadata. Compared with expert computation and token communication, this cost is negligible in practice.
\textcolor{red}{We do not attempt to solve the globally optimal hardware-aware routing problem online, since that would couple token-level expert selection with transient node loads, link occupancy, and future routing results. Instead, hardware-aware gating incorporates the marginal-cost terms into the expert-routing scores, making online selection deployable while keeping perturbation concentrated on low-impact routing changes.}

\subsubsection{Computation Penalty}

The computation penalty estimates the marginal increase in the maximum node compute load after selecting each candidate expert. Let $\mathbf{c}_{E\times 1\text{, init}}$ denote the base compute cost vector:
\begin{align}
&\mathbf{c}_{E\times 1\text{, init}} =(c_{0\text{, init}},c_{1\text{, init}},...,c_{E-1\text{, init}})^T,
\end{align}
where
\begin{align}
&c_{i\text{, init}}=3h \cdot IS \cdot \sum_{g \in G} \mathbb{I}_{E}(i \in g)
\end{align}
uses the same per-token cost as Eq.~(1). The initial per-node compute load is
\begin{align}
&\mathbf{C}_{1\times D\text{, init}} = \frac{\sum_{i=0}^{E-1} \mathbf{P}_{i} \cdot c_{i\text{, init}}}{\text{comp}}
\end{align}
where $\text{comp}$ is the compute throughput. If candidate expert $j$ is additionally selected, the updated expert-cost matrix is
\begin{align}
\mathbf{c}_{E\times E,\mathrm{incr}}
=
\mathbf{1}_E \mathbf{c}_{E\times 1,\mathrm{init}}^T
+ 3h \cdot IS \cdot I_E .
\end{align}
and the corresponding per-node load matrix is
\begin{align}
&\mathbf{C}_{E\times D\text{, incr}} = \frac{ c_{E\times E\text{, incr}}\cdot \mathbf{P}_{E\times D}  }{\text{comp}}
\end{align}
Thus,
\begin{align}
&\mathbf{T}_{\text{comp}} = \max_{0 \leq c \leq D-1} \{\mathbf{C}_{c,\text{incr}}\} - \max \{\mathbf{C}_{\text{init}}\} \cdot \mathbf{1}_{E}
\end{align}
penalizes experts that would increase the maximum node load.

\subsubsection{Communication Penalty}
The communication penalty uses deployment dispersion as a lightweight proxy for dispatch and aggregation cost:
\begin{align}
&T_i=\sum_{c=0}^{D-1} \mathbb{I}(P_{ic} \neq 0)
\end{align}
and
\begin{align}
&\mathbf{T}_{\text{comm}} = (T_0,T_1,...,T_{E-1})
\end{align}

\begin{figure}[!tb]
\centerline{\includegraphics[width=\linewidth]{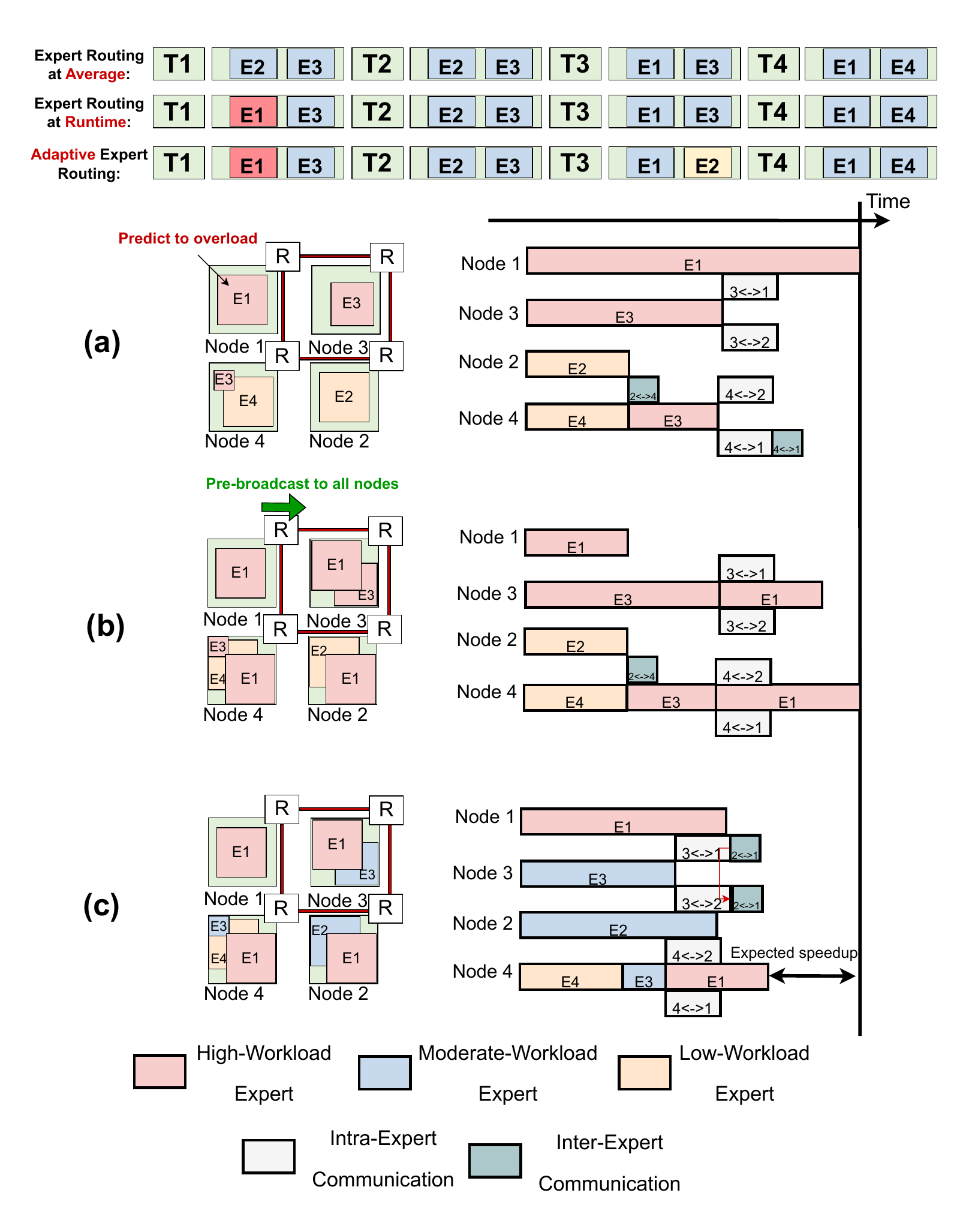}}
\caption{(a) Computation load detection, (b) Pre-broadcast without hardware-aware gating, (c) Pre-broadcast with hardware-aware gating.}
\label{fig:adaptive}
 \end{figure}

Fig.~\ref{fig:adaptive} illustrates hardware-aware gating. When scheduling can only pre-broadcast E1, conventional routing may shift the bottleneck from Node 1 to Nodes 3 and 4 through heavily loaded E3. Hardware-aware gating penalizes this high-cost boundary choice and redirects low-impact tokens toward available experts on less loaded nodes, reducing pressure on the hotspot path.

%% file: docs/5-experiment.tex
\section{Experimental Results}

\subsection{Experimental Setup}

\subsubsection{Models}
% We evaluate the performance of our proposed approach using three MoE models: Mixtral-8x7B-Instruct \cite{jiang2024mixtral} (mixtral), DeepSeek-V2-Lite-Chat \cite{deepseekv2} (deepseek) and Qwen2-57B-A14B-Instruct \cite{yang2024qwen2technicalreport} (qwen). All the models are large-scale MoE architectures that benefit from expert parallelism and tensor parallelism, and they are deployed on 3D NMP architectures with different mesh sizes and hardware configurations. The key parameters for both models are summarized in Table \ref{model}.

We evaluate \method~ on Mixtral-8x7B-Instruct-v0.1~\cite{jiang2024mixtral} (Mixtral), DeepSeek-V2-Lite-Chat~\cite{deepseekv2} (DeepSeek), Qwen2-57B-A14B-Instruct~\cite{yang2024qwen2technicalreport} (Qwen2), and \textcolor{red}{Qwen3.5-35B-A3B~\cite{qwen35model} (Qwen3.5), whose 256 routed experts test larger expert spaces.} Table~\ref{model} lists their key parameters.

\begin{table}[htbp]
\centering
\caption{Key configurations of the evaluated MoE models.}

\begin{center}
\resizebox{\columnwidth}{!}{ % 调整宽度为 80% 文本宽度，高度自动缩放
\begin{tabular}{c|c|c|c|c}

\hline\hline
\text{Configuration} & \text{Mixtral} & \text{DeepSeek} & \text{Qwen2} & \textcolor{red}{\text{Qwen3.5}} \\
\hline
\text{Number of Experts} & 8 & 64 & 64 & \textcolor{red}{256} \\
\hline
\text{Activated Experts per Token} & 2 & 6 & 8 & \textcolor{red}{8} \\
\hline
\text{Number of Layers} & 32 & 27 & 28 & \textcolor{red}{40} \\
\hline
\text{Hidden Size} & 4096 & 2048 & 3584 & \textcolor{red}{2048} \\
\hline
\text{Intermediate Size} & 14336 & 1408 & 2560 & \textcolor{red}{512} \\
\hline\hline
\end{tabular}}
\label{model}
\end{center}
\end{table}

\subsubsection{Baselines}
% The baseline deployment strategies include Tensor Parallelism (TP), Expert Parallelism (EP), and a hybrid TP-EP approach with compute balance. In the hybrid strategy, the 2D mesh is divided into sub-regions—8 for deepseek and qwen, and 2 for mixtral. Each sub-region applies EP, with TP used internally to parallelize expert computation. Experts are appropriately assigned to sub-regions to balance computation load, with each expert placed in only one sub-region. This strategy is widely used to mitigate load imbalance in large-scale systems.

% The baseline deployment strategies include Tensor Parallelism (TP), Expert Parallelism (EP), a hybrid TP-EP approach with compute balance, and the conference version of our work, HD-MoE~\cite{huang2025hd}. 
% In the hybrid TP-EP baseline, the 2D mesh is divided into sub-regions---8 for deepseek and qwen, and 2 for mixtral. Each sub-region applies EP, while TP is used within each sub-region to parallelize expert computation. Experts are assigned to sub-regions to balance computation load, and each expert is placed in only one sub-region. This strategy is widely used to mitigate load imbalance in large-scale systems.

Our baselines are TP, EP, a compute-balanced hybrid TP-EP scheme, and HD-MoE~\cite{huang2025hd}, the conference version of this work. 
The hybrid baseline partitions the mesh into regions---8 for DeepSeek and Qwen2, and 2 for Mixtral---uses EP across regions, TP within each region, and assigns each expert to one region for load balance.
HD-MoE includes hybrid placement and dynamic scheduling but not hardware-aware gating, isolating the benefit of adaptive routing.

% \subsubsection{Evaluation Metrics}
% We evaluate the performance of our approach and the baselines using the following metrics:

% \textbf{Normalized TBT (Time-Between-Tokens):} The latency between tokens during inference divided by that latency in Tensor Parallelism. 

% \textbf{MoE Decomposed Latency:} The time taken to process a batch of tokens in MoE layers, including both computation and communication time.
% \begin{itemize}
%     \item Computation Latency: The time spent on performing computations within each node.
%     \item Communication Latency: The time spent on transferring data between nodes.
% \end{itemize}

\subsubsection{Evaluation Metrics}
We report normalized TBT (Time-Between-Token, inter-token latency normalized to TP) and MoE latency breakdown, separating node-local computation from inter-node communication.

\subsubsection{Dataset}
% We use the MT Bench dataset \cite{10.5555/3666122.3668142} for evaluation, which is a widely adopted benchmark for LLMs, designed to measure the performance of LLMs on various tasks.
We use MT-Bench~\cite{10.5555/3666122.3668142} for latency under conversational workloads, and GSM8K~\cite{cobbe2021training}, HumanEval~\cite{chen2021evaluating}, ARC-E, and ARC-C~\cite{clark2018think} for model quality.

% \subsubsection{Offline Optimization}
% Our \textbf{Optimal Placement Strategy Searching} process typically takes only \textbf{several hours} for the entire procedure, which is considered relatively short for offline optimization.

\subsubsection{Hardware-aware Gating Parameters}
Hardware-aware gating uses computation and communication penalties $r_{\text{comp}}$ and $r_{\text{comm}}$. \textcolor{red}{We tune them per model and compute/bandwidth setting for the quality--speed trade-off, and reuse them across topologies with the same hardware setting because topology does not change the gating-score scale.} Table~\ref{tab:gating_params} lists the final settings. The two magnitudes are not comparable because $T_{\text{comp}}$ and $T_{\text{comm}}$ have different model-dependent scales.

\begin{table}[t]
\centering
\caption{Selected hyperparameters for hardware-aware gating. Each entry is $(r_{\mathrm{comp}},r_{\mathrm{comm}})$.}
\label{tab:gating_params}
\setlength{\tabcolsep}{2.5pt}
\resizebox{\columnwidth}{!}{%
\begin{tabular}{lccc}
\hline\hline
Model & \textcolor{red}{2.5 TFLOPS / 75 GB/s} & \textcolor{red}{5 TFLOPS / 50 GB/s} & \textcolor{red}{10 TFLOPS / 25 GB/s} \\
\hline
\textcolor{red}{Mixtral} & \textcolor{red}{$(-3.2{\times}10^4,-2.5{\times}10^{-2})$} & \textcolor{red}{$(-6.5{\times}10^4,-2.0{\times}10^{-2})$} & \textcolor{red}{$(-1.16{\times}10^5,-3.0{\times}10^{-2})$} \\
\textcolor{red}{DeepSeek} & \textcolor{red}{$(-3.5{\times}10^4,-5.0{\times}10^{-3})$} & \textcolor{red}{$(-6.8{\times}10^4,-2.2{\times}10^{-2})$} & \textcolor{red}{$(-1.05{\times}10^5,-3.5{\times}10^{-2})$} \\
\textcolor{red}{Qwen2} & \textcolor{red}{$(-1.2{\times}10^4,-5.0{\times}10^{-5})$} & \textcolor{red}{$(-1.8{\times}10^4,-1.0{\times}10^{-4})$} & \textcolor{red}{$(-7.0{\times}10^4,-4.0{\times}10^{-4})$} \\
\textcolor{red}{Qwen3.5} & \textcolor{red}{$(-1.0{\times}10^5,-2.0{\times}10^{-4})$} & \textcolor{red}{$(-2.0{\times}10^5,-1.0{\times}10^{-3})$} & \textcolor{red}{$(-2.0{\times}10^5,-1.0{\times}10^{-3})$} \\
\hline\hline
\end{tabular}
}
\end{table}

\subsubsection{Implementation and Offline Optimization}
We use an in-house end-to-end simulator that integrates expert computation, memory access, and inter-node communication; the latter uses the discrete-event NoC simulator in Sec.~\ref{comm_model}. \textcolor{red}{Table~\ref{tab:noc-sim-config} summarizes the NoC and hardware settings.}

\begin{table}[t]
\begingroup
\captionsetup{labelfont={color=red},textfont={color=red}}
\color{red}
\centering
\caption{Discrete-event NoC simulator configuration used in evaluation.}
\label{tab:noc-sim-config}
\setlength{\tabcolsep}{4pt}
\footnotesize
\begin{tabular}{@{}p{0.31\columnwidth}p{0.62\columnwidth}@{}}
\hline\hline
Parameter & Setting \\
\hline
Endpoint & Configurable number of NMP endpoints, each with local DRAM and compute \\
Topology & Configurable Mesh/Torus/Fat-tree; evaluated with $4\times4$, $4\times8$, $8\times8$ Mesh, $4\times8$ Torus, and 32-endpoint Fat-tree \\
Bandwidth & Configurable per-link bandwidth; evaluated with 25/50/75 GB/s and 50 GB/s Fat-tree tiers \\
Routing & XY for Mesh, minimum-hop wrap-around for Torus, up-down for Fat-tree \\
Traffic source & Token--expert routing traces; each event transfers expert-output bytes \\
Contention model & Hop-by-hop scheduling with per-directed-link FIFO serialization \\
\hline\hline
\end{tabular}
\endgroup
\end{table}

\textcolor{red}{The offline search (Sec.~\ref{sec:hybrid-parallelism}) typically finishes within several hours. Table~\ref{tab:offline_convergence} reports its cost across expert scales. Qwen2 takes longer than Qwen3.5 because its co-activations span more distinct expert groups, increasing group-placement variables; BO uses the same 70-iteration Link Balance setting as the topology experiments.}

\begin{table}[t]
\begingroup
\captionsetup{labelfont={color=red},textfont={color=red}}
\color{red}
\centering
\caption{Offline optimization cost and LP characteristics under 5 TFLOPS / 50 GB/s, (4,8). X vars.=$2ED$ expert-placement variables; $G$ is the average number of co-activated expert groups; Y vars.=$\bar{G}D$; Top-100 denotes
the frequency mass covered by the 100 most frequent groups; $H_G$
denotes normalized group-frequency entropy; LP reports average
solving time.}
\label{tab:offline_convergence}
\setlength{\tabcolsep}{2pt}
\footnotesize
\makebox[\columnwidth][c]{%
\resizebox{0.9\columnwidth}{!}{
\begin{tabular}{@{}cccccccccc@{}}
\hline\hline
Model & Exp. & MoE Lay. & X vars. & $G$ & Y vars. & Top-100 & $H_G$ & LP (s) & BO (min) \\
\hline
Mixtral & 8 & 32 & 512 & 27.9 & 0.9K & 100.0\% & 0.908 & 1.95 & 49.61 \\
DeepSeek & 64 & 26 & 4.1K & 2030.7 & 65.0K & 23.46\% & 0.929 & 160.22 & 43.69 \\
Qwen2 & 64 & 28 & 4.1K & 4987.0 & 159.6K & 13.27\% & 0.961 & 5205.78 & 58.23 \\
Qwen3.5 & 256 & 40 & 16.4K & 4886.9 & 156.4K & 37.75\% & 0.811 & 550.36 & 50.29 \\
\hline\hline
\end{tabular}}}
\endgroup
\end{table}

\begin{figure}[t]
    \centering
    \includegraphics[width=\linewidth]{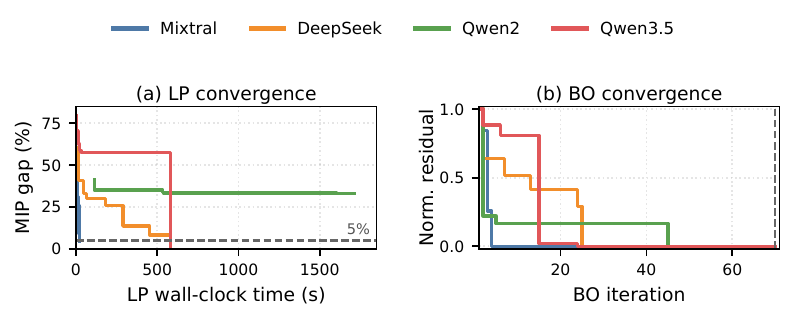}
    \caption{\textcolor{red}{Offline-optimization convergence. (a) LP optimality gap over time. (b) Link Balance objective improvement during BO.}}
    \label{fig:offline-convergence}
\end{figure}

\textcolor{red}{Fig.~\ref{fig:offline-convergence} further shows convergence: LP finishes within minutes for Mixtral, DeepSeek, and Qwen3.5, while Qwen2 needs a longer search; BO becomes stable within about 45 iterations under the 70-iteration budget.}

\subsection{End-to-End Performance}

% We evaluate the end-to-end performance of our method across different hardware configurations and 2D mesh sizes inferred in different batch sizes. The experiments are conducted using three hardware configurations, each with varying compute throughput and communication bandwidth:
We evaluate end-to-end performance under three compute/bandwidth settings: 2.5 TFLOPS/75 GB/s, 5 TFLOPS/50 GB/s, and 10 TFLOPS/25 GB/s.

% Additionally, we compare the performance across three different 2D mesh sizes: (4,4), (4,8), and (8,8), with 5 TFLOPS compute throughput and 50 GB/s bandwidth for consistency.
We also evaluate $4\times4$, $4\times8$, and $8\times8$ meshes at 5 TFLOPS/50 GB/s. 
\textcolor{red}{Under the same 32-endpoint, 50 GB/s setting, we compare Mesh ($4\times8$), Torus with wrap-around routing, and a 4-pod full-bisection Fat-tree with up-down routing and $\text{BW}_{\text{el}}=\text{BW}_{\text{la}}=\text{BW}_{\text{ac}}=50$ GB/s.}
Each node has 625 GB/s memory bandwidth. \textcolor{red}{Table~\ref{tab:mem-cap} lists total and per-node DRAM capacity limits.}

\begin{table}[!tb]
\begingroup
\captionsetup{labelfont={color=red},textfont={color=red}}
\color{red}
\centering
\caption{DRAM capacity limits used in evaluation.}
\label{tab:mem-cap}
\setlength{\tabcolsep}{2.5pt}
\footnotesize
\resizebox{\columnwidth}{!}{
\begin{tabular}{lccccc}
\hline\hline
Model & $\rho_{\text{mem}}$ & Total limit & \multicolumn{3}{c}{Per-node limit} \\
\cline{4-6}
 & & & 4$\times$4 & 4$\times$8 & 8$\times$8 \\
\hline
Mixtral & 1.10 & 99.2 GB & 6.20 GB & 3.10 GB & 1.55 GB \\
DeepSeek & 1.25 & 37.4 GB & 2.34 GB & 1.17 GB & 0.58 GB \\
Qwen2 & 1.25 & 123.4 GB & 7.71 GB & 3.85 GB & 1.93 GB \\
Qwen3.5 & 1.25 & 80.5 GB & 5.03 GB & 2.52 GB & 1.26 GB \\
\hline\hline
\end{tabular}}
\endgroup
\end{table}

\begin{figure*}[!tb]
    \centering
    % \scriptsize
    \begin{tabular}{ccccc}
    
        \textcolor{color1}{\rule{1em}{1em}} TP & 
        \textcolor{color2}{\rule{1em}{1em}} EP & 
        \textcolor{color3}{\rule{1em}{1em}} Compute Balance & 
        \textcolor{color4}{\rule{1em}{1em}} HD-MoE &
        \textcolor{color5}{\rule{1em}{1em}} HDA-MoE\\
    \end{tabular} \\[0.5em]
 
\centerline{\includegraphics[width=\linewidth]{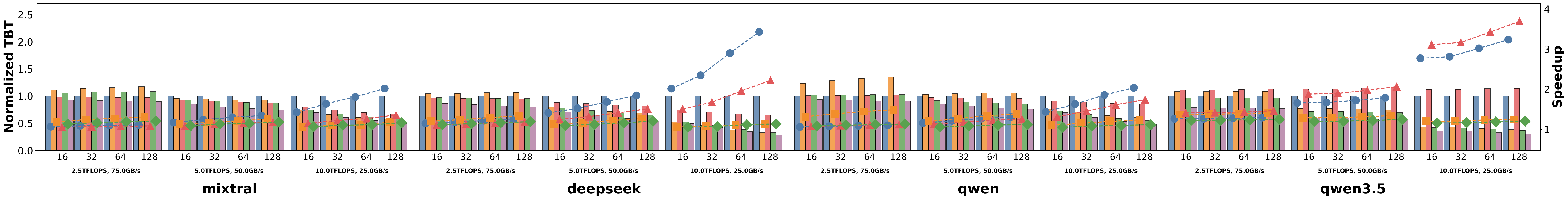}}
\caption{End-to-end performance under different hardware configurations}
\label{config}
\end{figure*}

\textbf{Better TBT latency through different hardware configurations:} 
% The results of these experiments, shown in Fig. \ref{config} and Fig. \ref{mesh}. Results in Fig. \ref{config} reveal how different methods respond to shifts in compute-to-communication ratios. When computation is limited and communication bandwidth is abundant (2.5 TFLOPS, 75 GB/s), EP suffers from severe workload imbalance, resulting in suboptimal TBT latency. In contrast, when computation is sufficient but communication becomes a bottleneck (10 TFLOPS, 25 GB/s), TP incurs heavy all-reduce communication costs, leading to degraded latency performance. Additionally, worth noting is that, for qwen, the expert routing exhibits high imbalance (Fig.~\ref{fig:motivation}(a)), causing significant overhead in EP.
Fig.~\ref{config} compares compute/bandwidth settings. EP suffers from load imbalance at 2.5 TFLOPS/75 GB/s, while TP suffers from all-reduce overhead at 10 TFLOPS/25 GB/s, especially on the highly imbalanced Qwen2 routing in Fig.~\ref{fig:motivation}(a). The compute-balanced hybrid improves load balance but ignores topology; HD-MoE adds joint placement and dynamic scheduling; HDA-MoE further improves results through adaptive routing.

% The Hybrid TP-EP with Compute-Balanced baseline achieves better performance by distributing expert load more evenly, but ignores communication topology, leading to degraded performance under constrained bandwidth.

% In contrast, our Node-Link Balance strategy jointly considers both computation and communication during expert placement. It minimizes per-node compute load, inter-node communication volume, and per-link congestion. As a result, it consistently outperforms all baselines across different system configurations.

% On average, our method achieves a speedup ranging from \textbf{1.1$\times$ to 1.8$\times$} compared to TP, \textbf{1.1$\times$ to 1.5$\times$} compared to EP, and \textbf{1.0$\times$ to 1.4$\times$} compared to Hybrid TP-EP with Compute-Balanced.

\begin{figure*}[!tb]
    \centering
\centerline{\includegraphics[width=\linewidth]{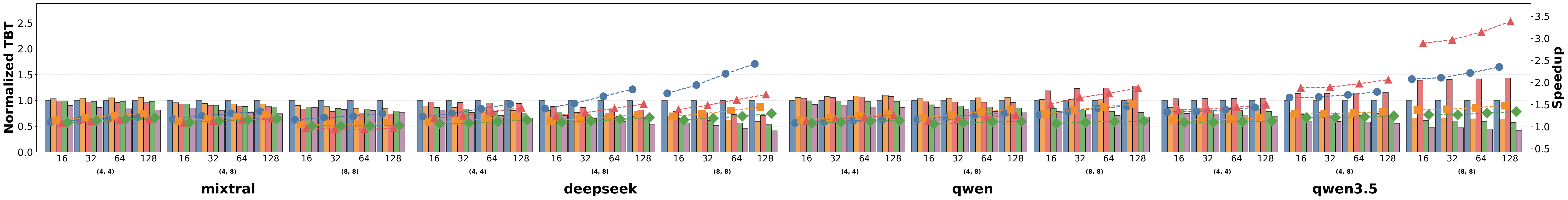}}
\caption{End-to-end performance under different mesh sizes}
\label{mesh}
\end{figure*}

\textbf{Better TBT latency through different mesh size:} 
% We evaluate the impact of mesh topology on TBT latency under a fixed configuration (5 TFLOPS, 50 GB/s). As shown in Fig. \ref{mesh}, our Node-Link Balance strategy consistently achieves low latency across mesh sizes, demonstrating strong adaptability.
Fig.~\ref{mesh} shows that \method~maintains low TBT across mesh sizes, indicating adaptability to topology scaling.

\begin{figure*}[!tb]
    \centering
    \includegraphics[width=\linewidth]{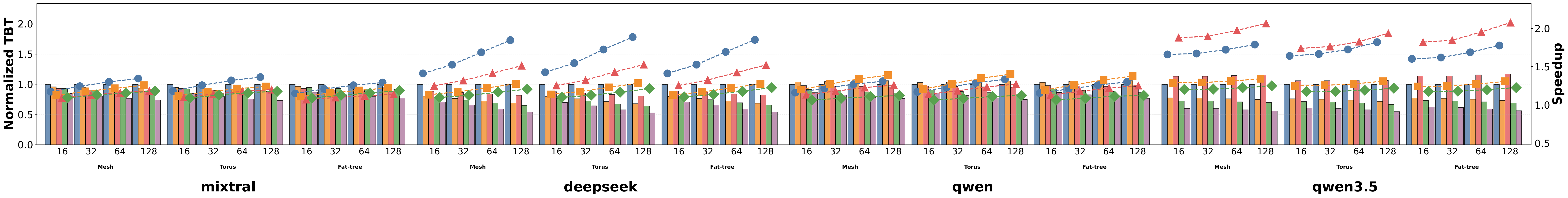}
    \caption{\textcolor{red}{End-to-end performance under different topologies}}
    \label{fig:topology}
\end{figure*}

\textcolor{red}{\textbf{Better TBT latency through alternative topologies:}
Fig.~\ref{fig:topology} shows that HDA-MoE outperforms all baselines on Torus and Fat-tree, indicating that Link Balance generalizes beyond XY-mesh routing.}

Overall, HDA-MoE achieves 1.1$\times$--\textcolor{red}{3.4}$\times$ over TP, 1.1$\times$--\textcolor{red}{1.5}$\times$ over EP, 1.1$\times$--\textcolor{red}{3.7}$\times$ over the Hybrid TP-EP compute-balanced baseline, and \textcolor{red}{1.1}$\times$--\textcolor{red}{1.3}$\times$ over HD-MoE.

% An exception occurs in mixtral with an (8,8) mesh, where the Hybrid TP-EP with Compute-Balanced baseline achieves slightly better latency. This is likely due to mixtral’s small number of experts, which must be spread across multiple nodes. In large mesh topologies, where communication regularity is more critical, the hybrid baseline benefits from its structured TP communication and moderate message volume.

% Overall, our method remains effective across models and mesh sizes, particularly when the number of experts and the topology scale are well matched.
% One exception appears for mixtral on the (8,8) mesh, where the compute-balanced hybrid baseline performs slightly better.
% This is likely due to mixtral’s small number of experts, which must be spread across multiple nodes. In large mesh topologies, where communication regularity is more critical, the regular communication pattern of the hybrid baseline becomes more favorable. 
% Overall, \method~remains effective across models and mesh sizes, especially when the expert count better matches the topology scale.

\subsection{Latency Breakdown}

\begin{figure}[!tb]
    \centering
    % \scriptsize
% \begin{tabular}{c c c}
%     \multicolumn{3}{c}{
%         \textcolor{color1}{\rule{1em}{1em}} MoE Compute \quad
%         \textcolor{color2}{\rule{1em}{1em}} Communication
%     } \\

%     \textcolor{color3}{\rule{1em}{1em}} MoE Memory Access &
%     \textcolor{color4}{\rule{1em}{1em}} Attn Compute &
%     \textcolor{color5}{\rule{1em}{1em}} Attn Memory Access
% \end{tabular}
\setlength{\tabcolsep}{3pt}
\resizebox{\columnwidth}{!}{
\begin{tabular}{c c c}
    \multicolumn{3}{c}{
        \textcolor{color1}{\rule{1em}{1em}} MoE Compute \quad
        \textcolor{color2}{\rule{1em}{1em}} Communication
    } \\

    \textcolor{color3}{\rule{1em}{1em}} MoE Memory Access &
    \textcolor{color4}{\rule{1em}{1em}} Attn Compute &
    \textcolor{color5}{\rule{1em}{1em}} Attn Memory Access
\end{tabular}
}
\setlength{\tabcolsep}{6pt}
\centerline{\includegraphics[width=\linewidth]{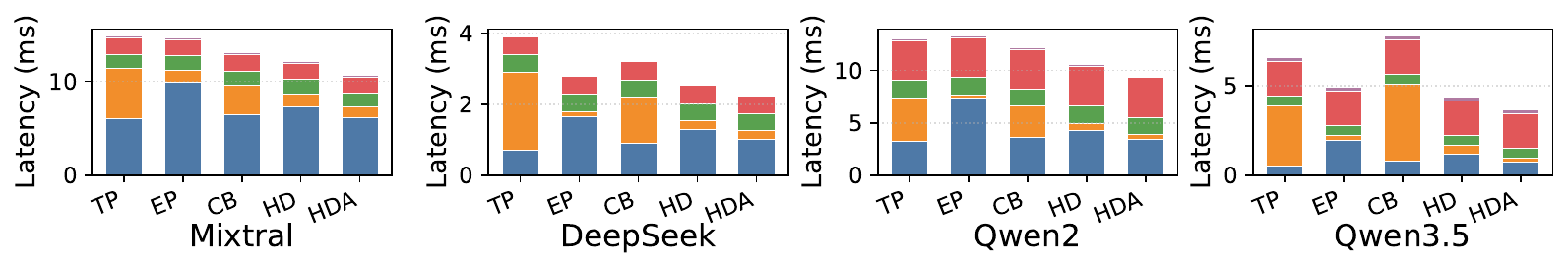}}
\caption{End-to-end latency breakdown. CB denotes compute-balanced TP-EP; HD denotes HD-MoE; HDA denotes HDA-MoE. Attn includes attention layers and shared experts.}
\label{breakdown}
\end{figure}

Fig.~\ref{breakdown} decomposes MoE-layer latency into computation, communication, and memory access under 5 TFLOPS compute, 50 GB/s communication, and 625 GB/s memory access.

Compared with TP, \method~reduces all-reduce communication through better placement and locality; compared with EP, it lowers compute tail latency by alleviating expert imbalance. It further improves over HD-MoE through hardware-aware gating.

Memory access is a small latency component, so the remaining bottlenecks are computation and communication, which \method~targets through placement, scheduling, and routing.

\subsection{Accuracy Evaluation}

Since hardware-aware gating modifies expert selection, Table~\ref{tab:accuracy} compares original and HAG routing. HAG keeps comparable accuracy, so the efficiency gain does not come from model-quality degradation.
% \begin{table}[t]
% \centering
% \caption{Accuracy comparison between the default routing strategy and our hardware-aware gating method on representative benchmarks. For \method, each entry is reported as ``a/b/c'', corresponding to the three hardware configurations: (10 TFLOPS, 25 GB/s), (5.0 TFLOPS, 50 GB/s), (2.5 TFLOPS, 75 GB/s), and respectively.}
% \label{tab:accuracy}
% % \scriptsize
% \setlength{\tabcolsep}{2pt} % 默认是 6pt
% \centering
% \begin{tabular}{l l  c c c c }
% \hline\hline
% \text{Model}    & Method   & GSM8K & HumanEval & ARC-E & ARC-C \\

% % &&Baseline&\method&Baseline&\method&Baseline&\method&Baseline&\method \\
% \hline
% \text{mixtral}    & Baseline &  48.1 &    42.1   & 85.9  &80.9 \\
%          & \method    &    64.1/64.1/64.2   &  42.1/42.1/42.1  &  88.3/88.3/88.4  &  80.6/80.6/80.6   \\

% \hline
% \text{deepseek}  & Baseline & 70.1 & 43.3   & 68.2  &56.5 \\
%          & \method    &   69.6/70.5/69.9     & 43.3/43.9/45.1   &  69.5/67.5/67.9  &  56.9/56.5/57.2     \\

% \hline
% \text{qwen}  & Baseline & 78.2  &    50.0   & 93.2 & 89.3\\
%          & \method    &  78.2/78.0/78.0     & 51.2/50.0/50.0    &  93.2/93.0/93.2 &    89.0/89.3/89.3  \\

% \hline\hline
% \end{tabular}
% \end{table}

\begin{table}[t]
\centering
\caption{\textcolor{red}{Accuracy on representative benchmarks. HAG entries are ``a/b/c'' for (10 TFLOPS, 25 GB/s), (5 TFLOPS, 50 GB/s), and (2.5 TFLOPS, 75 GB/s).}}
\label{tab:accuracy}

\setlength{\tabcolsep}{2.5pt}
\renewcommand{\arraystretch}{1.08}
\footnotesize
\resizebox{\columnwidth}{!}{%
\begin{tabular}{@{}l @{\hspace{4pt}} l c c c c@{}}
\hline\hline
\textcolor{red}{Model} & \textcolor{red}{Routing} & GSM8K & HumanEval & ARC-E & ARC-C \\
\hline

\multirow{2}{*}{\textcolor{red}{Mixtral}} & \textcolor{red}{Original} & 63.2 & 42.1 & 85.9 & 80.9 \\
 & \textcolor{red}{HAG} & \textcolor{red}{64.1/64.1/64.2} & \textcolor{red}{45.1/43.9/45.1} & \textcolor{red}{85.6/86.3/86.1} & \textcolor{red}{82.6/83.3/81.3} \\

\hline
\multirow{2}{*}{\textcolor{red}{DeepSeek}} & \textcolor{red}{Original} & 70.1 & 43.3 & 68.2 & 56.5 \\
 & \textcolor{red}{HAG} & \textcolor{red}{70.6/70.7/70.1} & \textcolor{red}{42.7/44.5/43.3} & \textcolor{red}{69.5/68.9/68.9} & \textcolor{red}{57.5/57.5/58.5} \\

\hline
\multirow{2}{*}{\textcolor{red}{Qwen2}} & \textcolor{red}{Original} & 78.2 & 50.0 & 93.2 & 89.3 \\
 & \textcolor{red}{HAG} & \textcolor{red}{77.9/77.6/78.4} & \textcolor{red}{50.6/50.6/49.4} & \textcolor{red}{93.2/93.3/93.3} & \textcolor{red}{89.0/89.0/89.3} \\

\hline
\multirow{2}{*}{\textcolor{red}{Qwen3.5}} & \textcolor{red}{Original} & \textcolor{red}{74.8} & \textcolor{red}{65.9} & \textcolor{red}{93.3} & \textcolor{red}{90.0} \\
 & \textcolor{red}{HAG} & \textcolor{red}{74.5/74.7/74.5} & \textcolor{red}{65.2/64.0/65.2} & \textcolor{red}{93.3/93.5/94.0} & \textcolor{red}{89.3/90.3/91.0} \\

\hline\hline
\end{tabular}%
}

\end{table}

\subsection{Routing Fidelity and Expert Substitutability Analysis}
\label{sec:substitutability}

\textcolor{red}{We next analyze boundary expert substitutability and end-to-end prediction perturbation under hardware-aware routing.}

\begin{figure}[!tb]
    \centering
    \includegraphics[width=\linewidth]{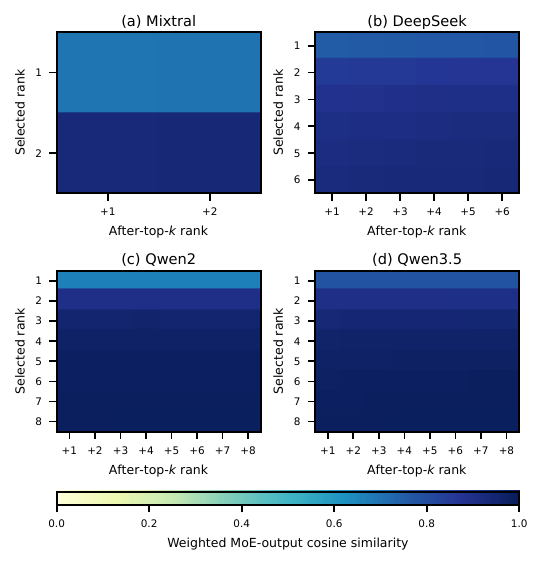}
    \caption{\textcolor{red}{Output similarity under boundary expert substitution. Entry $(i,j)$ is the cosine similarity after replacing the $i$-th selected expert with the $(k{+}j)$-th ranked expert. Rows index the selected rank $i$; columns index the after-top-$k$ candidate rank $j$.}}
    \label{fig:rankwise-replacement}
\end{figure}

\textcolor{red}{\textbf{Boundary expert substitutability.}
For each token and layer, we replace the $i$-th selected top-$k$ expert with the expert at overall rank $k{+}j$ (the $j$-th after-top-$k$ candidate), reuse the candidate's original gating weight, and measure the cosine similarity between the original and substituted score-weighted top-$k$ outputs. Fig.~\ref{fig:rankwise-replacement} reports the layer- and dataset-averaged similarity for each $(i,j)$ pair.}

\textcolor{red}{Across the four models, non-Top-1 replacements achieve an average similarity of 0.949, showing that lower-ranked boundary experts are highly substitutable. In contrast, Top-1 replacements are more sensitive, so the practical perturbation of hardware-aware gating should be assessed together with Top-1 routing retention.}

\textcolor{red}{\textbf{End-to-end routing impact and model perturbation.}
We report KL between the next-token probability distributions produced by original and hardware-aware routing, PPL ratio between the two routing modes, Top-1 prediction agreement between their predicted tokens, and Top-1 routing retention, i.e., the fraction of tokens whose original highest-scored expert remains in the adjusted top-$k$ set. Metrics are computed on WikiText-103~\cite{merity2017pointer}; Cache-prior~\cite{skliar2025cache} is included as a routing-adjustment reference.}

\begin{table}[t]
\begingroup
\captionsetup{labelfont={color=red},textfont={color=red}}
\color{red}
\centering
\caption{Routing impact under routing adjustment. \method~entries are ``a/b/c'' for (10 TFLOPS, 25 GB/s), (5 TFLOPS, 50 GB/s), and (2.5 TFLOPS, 75 GB/s).}
\label{tab:routing-perturbation}
\setlength{\tabcolsep}{3pt}
\resizebox{\columnwidth}{!}{
\begin{tabular}{@{}llcccc@{}}
\hline\hline
Model & Routing & KL $\downarrow$ & PPL ratio $\downarrow$ & Top-1 rout. ret. $\uparrow$ & Top-1 pred. agr. $\uparrow$ \\
\hline
Mixtral & Cache-prior & 0.0526 & 1.0403 & 100.00\% & 91.32\% \\
        & \method & \textcolor{red}{0.0882/0.0935/0.0996} & \textcolor{red}{1.0769/1.0907/1.0923} & \textcolor{red}{85.00/85.00/83.75\%} & \textcolor{red}{88.43/88.43/87.99\%} \\
\hline
DeepSeek & Cache-prior & 0.1044 & 1.0333 & 99.38\% & 85.69\% \\
        & \method & \textcolor{red}{0.1820/0.1812/0.1504} & \textcolor{red}{1.0774/1.0841/1.0627} & \textcolor{red}{95.00/97.50/96.25\%} & \textcolor{red}{81.76/82.01/84.09\%} \\
\hline
Qwen2 & Cache-prior & 0.1010 & 1.0360 & 95.63\% & 85.86\% \\
        & \method & \textcolor{red}{0.1248/0.0593/0.0919} & \textcolor{red}{1.0771/1.0365/1.0391} & \textcolor{red}{76.88/82.50/81.25\%} & \textcolor{red}{85.49/90.00/87.62\%} \\
\hline
Qwen3.5 & Cache-prior & 0.4435 & 1.5721 & 100.00\% & 72.38\% \\
        & \method & \textcolor{red}{0.2423/0.4148/0.1159} & \textcolor{red}{1.2757/1.5611/1.1147} & \textcolor{red}{77.50/73.75/93.75\%} & \textcolor{red}{79.56/73.04/86.74\%} \\
\hline\hline
\end{tabular}}
\endgroup
\end{table}

\textcolor{red}{Table~\ref{tab:routing-perturbation} shows controlled perturbation across settings. Top-1 routing retention is generally high; together with Fig.~\ref{fig:rankwise-replacement}, this indicates that routing changes have limited impact because lower-ranked boundary experts are more substitutable than the most important expert. Expert redundancy nevertheless varies across models: Mixtral is more sensitive to boundary replacements, whereas DeepSeek, Qwen2, and Qwen3.5 admit smaller or comparable distributional shifts. Together with Table~\ref{tab:accuracy}, these results show that \method~exploits model-dependent boundary substitutability while largely preserving the original model function.}

\subsection{Ablation Study}

% We further conduct an ablation study focused on the contribution of the \textbf{Node balance}, \textbf{Link balance}, and \textbf{Dynamic scheduling optimization} for deepseek.
We conduct ablation studies on \textbf{Node Balance}, \textbf{Link Balance}, \textbf{Memory-Capacity Constraint}, \textbf{Dynamic Scheduling}, and \textbf{Hardware-aware Gating} using DeepSeek.

\subsubsection{Node Balancing}

\begin{figure}[!tb]
\centering

\resizebox{\columnwidth}{!}{
\begin{tabular}{cccc}
\textcolor{color1}{\rule{1em}{1em}} TP &
\textcolor{color2}{\rule{1em}{1em}} EP &
\textcolor{color3}{\rule{1em}{1em}} Compute Balance &
\textcolor{color4}{\rule{1em}{1em}} Node-Link Balance
\end{tabular}
}

\vspace{0.1em}

\includegraphics[width=\linewidth]{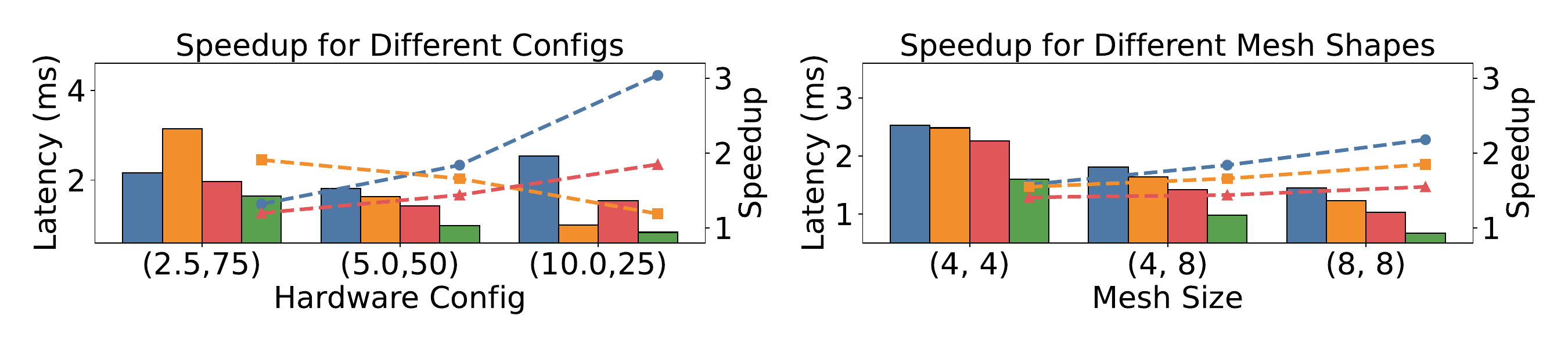}

\caption{Speedup of Node Balance on DeepSeek.}
\label{node}
\end{figure}

% We first evaluate the effect of the Node Balance stage in Fig. \ref{node}. It achieves \textbf{1.0$\times$ to 3.0$\times$} speedup over TP and EP, and \textbf{1.5$\times$} over Hybrid TP-EP across various configurations, by improving compute load balance (vs. EP) and reducing communication volume (vs. TP and hybrid).
Fig.~\ref{node} evaluates Node Balance. It achieves \textbf{1.0$\times$--3.0$\times$} speedup over TP/EP and \textbf{1.5$\times$} over compute-balanced hybrid by improving compute balance and reducing communication volume.

\begin{figure}[!tb]
    \centering
    % \scriptsize
    
    \begin{tabular}{cccc}
        \textcolor{color1}{\rule{1em}{1em}} EP & 
        \textcolor{color2}{\rule{1em}{1em}} Node-Link Balance & 

    \end{tabular}
\centerline{\includegraphics[width=\linewidth]{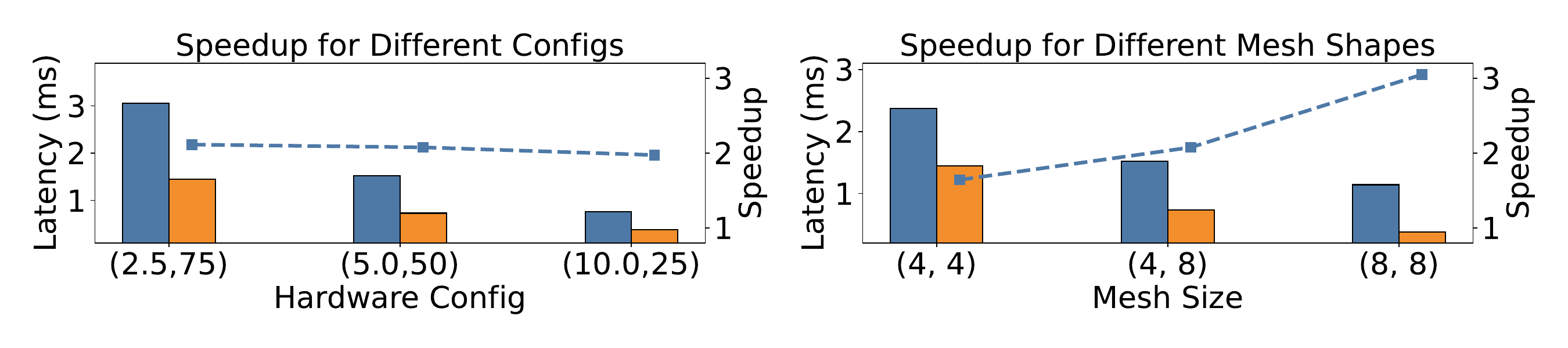}}
\caption{Computation-latency reduction from Node Balance in DeepSeek.}
\label{node_comp}
\end{figure}

% Next, we specifically examine Node Balance's effect on compute imbalance in EP, by measuring its impact on computation latency within MoE layers As shown in Fig. \ref{node_comp}, on average, Node Balance reduces EP's compute tail latency by \textbf{2.0$\times$}, confirming its effectiveness in mitigating the routing skew commonly observed in MoE models.
\textbf{Better computation latency:} Fig.~\ref{node_comp} further isolates Node Balance's impact on compute latency. It reduces EP's compute tail latency by \textbf{2.0$\times$} on average, confirming its effectiveness in mitigating MoE load skew.

\begin{figure}[!tb]
\centerline{\includegraphics[width=\linewidth]{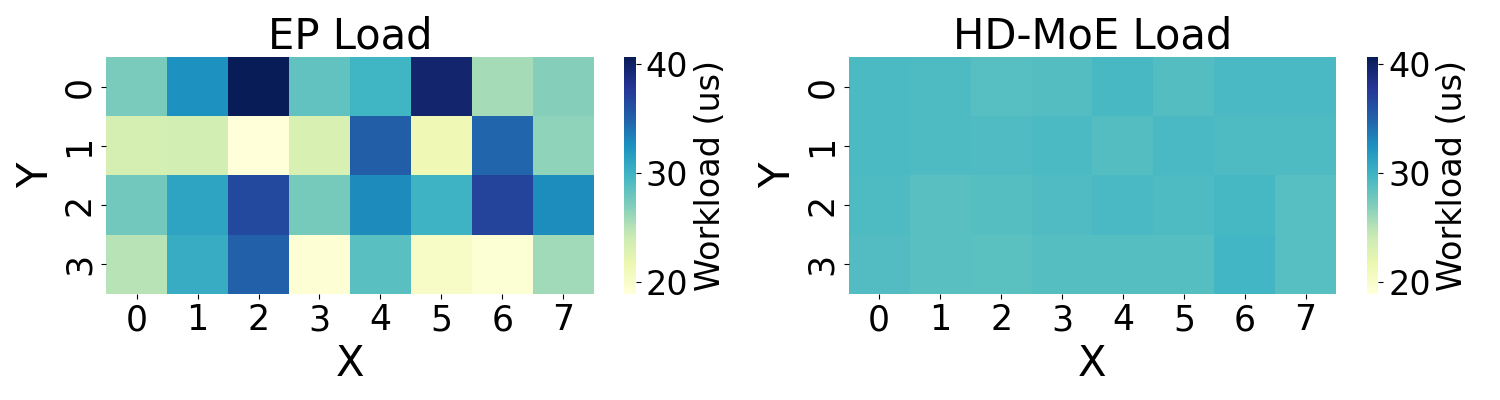}}
\caption{Node-level resource utilization before and after Node Balance.}
\label{load}
\end{figure}

% Fig. \ref{load} shows per-node compute and communication load before and after applying Node Balance. The optimized placement achieves noticeably better load balance than standard EP, which often exhibits severe hotspots.
\textbf{Better load balance:} Fig.~\ref{load} visualizes per-node compute and communication load before and after Node Balance. Compared with EP, the optimized placement produces a more balanced distribution and removes severe hotspots.

\subsubsection{Link Balancing}
\begin{figure}[!tb]
    \centering
    % \scriptsize
    \resizebox{\columnwidth}{!}{
    \begin{tabular}{cccc}
        \textcolor{color1}{\rule{1em}{1em}} TP & 
        \textcolor{color2}{\rule{1em}{1em}} EP & 
        \textcolor{color3}{\rule{1em}{1em}} Compute Balance & 
        \textcolor{color4}{\rule{1em}{1em}} Node-Link Balance\\
    \end{tabular}}
\vspace{0.1em}
\centerline{\includegraphics[width=\linewidth]{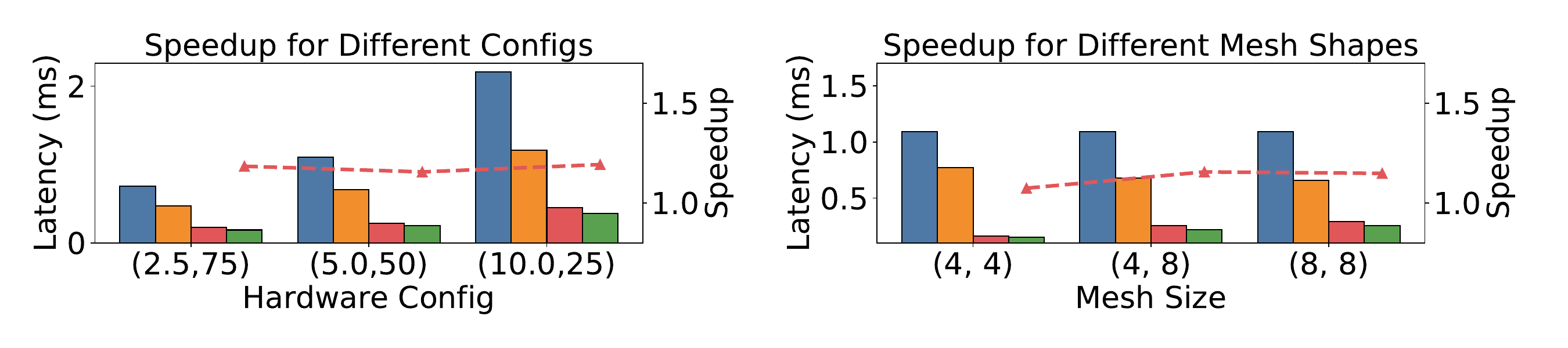}}
\caption{Communication benefit of Link Balance on DeepSeek.}
\label{link}
\end{figure}

% We further evaluate the contribution of the Link Balance stage by isolating its impact on communication latency in Fig. \ref{link}. Specifically, we compare against three baselines: TP, Hybrid TP-EP with Compute-Balanced, and the Node Balance without physical mapping optimization.
Fig.~\ref{link} evaluates the effect of Link Balance by comparing against TP, the hybrid baseline, and Node Balance without physical mapping optimization.

% Thanks to the Bayesian Optimization–based mapping strategy, Link Balance produces more communication-friendly physical mappings by assigning logical clusters to physical nodes in a topology-aware manner. This significantly reduces link congestion and results in lower communication latency than both TP and hybrid baselines, which rely on regular but heavy communication.
% Compared to the Node Balance–only deployment, Link Balance can also achieve an average \textbf{1.2$\times$} reduction in communication latency, highlighting the importance of mapping logical clusters to physical nodes with awareness of network structure.
% By assigning logical clusters to physical nodes in a topology-aware manner, Link Balance reduces link congestion and lowers communication latency relative to TP and the hybrid baseline.
\textbf{Better communication latency:} Topology-aware Link Balance reduces link congestion and communication latency compared with TP and the hybrid baseline. Compared with Node Balance alone, it achieves an average \textbf{1.2$\times$} communication-latency reduction.

\begin{figure}[!tb]
\centerline{\includegraphics[width=\linewidth]{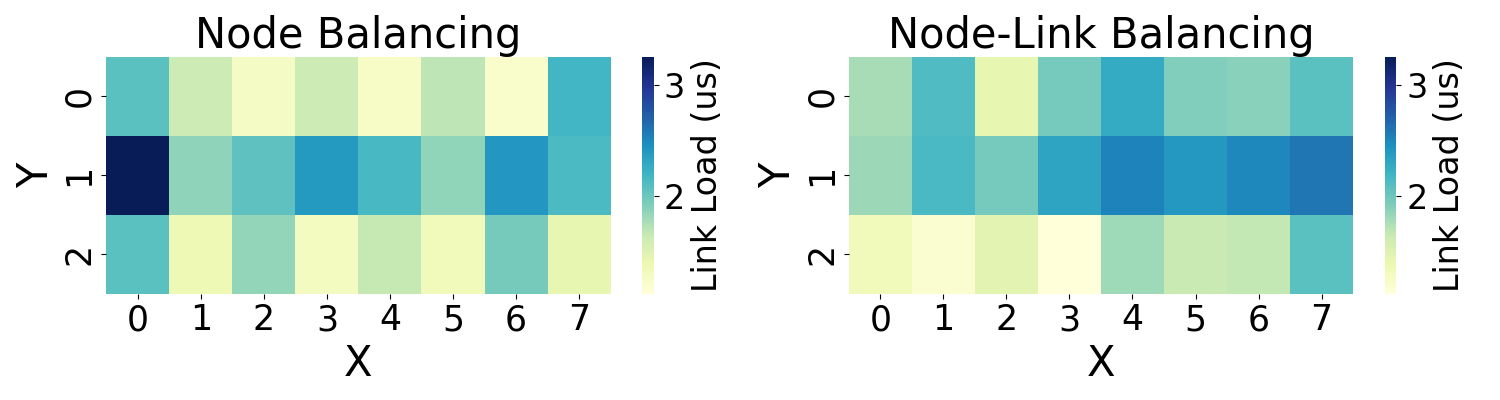}}
\caption{Link-level resource utilization before and after Link Balance.}
\label{link_eg}
\end{figure}

% Fig. \ref{link_eg} visualizes NoC link utilization using heatmaps. Compared to the Node Balance stage, the optimized placement after Link Balance leads to visibly more balanced link-level traffic, with less link congestion and better distribution across the mesh.
\textbf{Less link congestion:} Fig.~\ref{link_eg} further shows that Link Balance distributes traffic more evenly across the mesh, visibly reducing congestion compared with Node Balance alone.

\subsubsection{Memory-Capacity Constraint}

\begin{figure}[!tb]
    \centering
    \includegraphics[width=\linewidth]{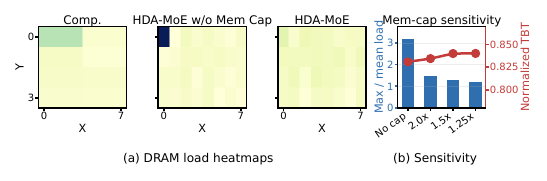}
    \caption{\textcolor{red}{Memory-capacity analysis. (a) Expert-weight storage across DRAM nodes. (b) DRAM-load ratio and normalized TBT when varying $\rho_{\text{mem}}$.}}
    \label{fig:memory-cap}
\end{figure}

\textcolor{red}{Fig.~\ref{fig:memory-cap} evaluates the memory-capacity constraint in Eq.~\eqref{eq:memcap} under 5 TFLOPS / 50 GB/s. Fig.~\ref{fig:memory-cap}(b) varies the cap factor $\rho_{\text{mem}}$ and reports normalized TBT from placements re-optimized with the corresponding constraint.}

\textcolor{red}{\textbf{Better memory-capacity balance:}
Without the memory-capacity constraint, HDA-MoE may concentrate expert weights on a few nodes; adding the constraint balances DRAM load more evenly. As $\rho_{\text{mem}}$ becomes tighter, the max/mean DRAM-load ratio decreases consistently, while normalized TBT remains close to the unconstrained case. Thus, \method~ improves memory-capacity utilization without sacrificing its end-to-end advantage.}

% \begin{figure}[!tb]
% \centering
% \scriptsize
%     \begin{tabular}{cccc}
%         \textcolor{color1}{\rule{1em}{1em}} Static Deployment & 
%         \textcolor{color2}{\rule{1em}{1em}} Dynamic Deployment & 

%     \end{tabular} \\[0.5em]
% \begin{subfigure}[t]{0.24\textwidth}
%     \includegraphics[width=\linewidth]{figs/dynamic_deployment_latency_speedup.pdf}
%     \caption{}
%     \label{dynamic2}
% \end{subfigure}
% \hfill
% \begin{subfigure}[t]{0.24\textwidth}
%     \includegraphics[width=\linewidth]{figs/dynamic_deployment_latency_speedup2.pdf}
%     \caption{}
%     \label{dynamic5}
% \end{subfigure}
% \caption{Static vs. dynamic placement under different inference scenarios. (a) 2 experts pre-broadcast, (b) 5 experts pre-broadcast.}
% \label{dynamic}
% \end{figure}
\begin{figure}[!tb]
\centering
% \scriptsize
\begin{tabular}{cc}
\textcolor{color1}{\rule{1em}{1em}} Static Deployment &
\textcolor{color2}{\rule{1em}{1em}} Dynamic Deployment
\end{tabular}

\vspace{0.1em}

\subfloat{
    \includegraphics[width=0.45\columnwidth]{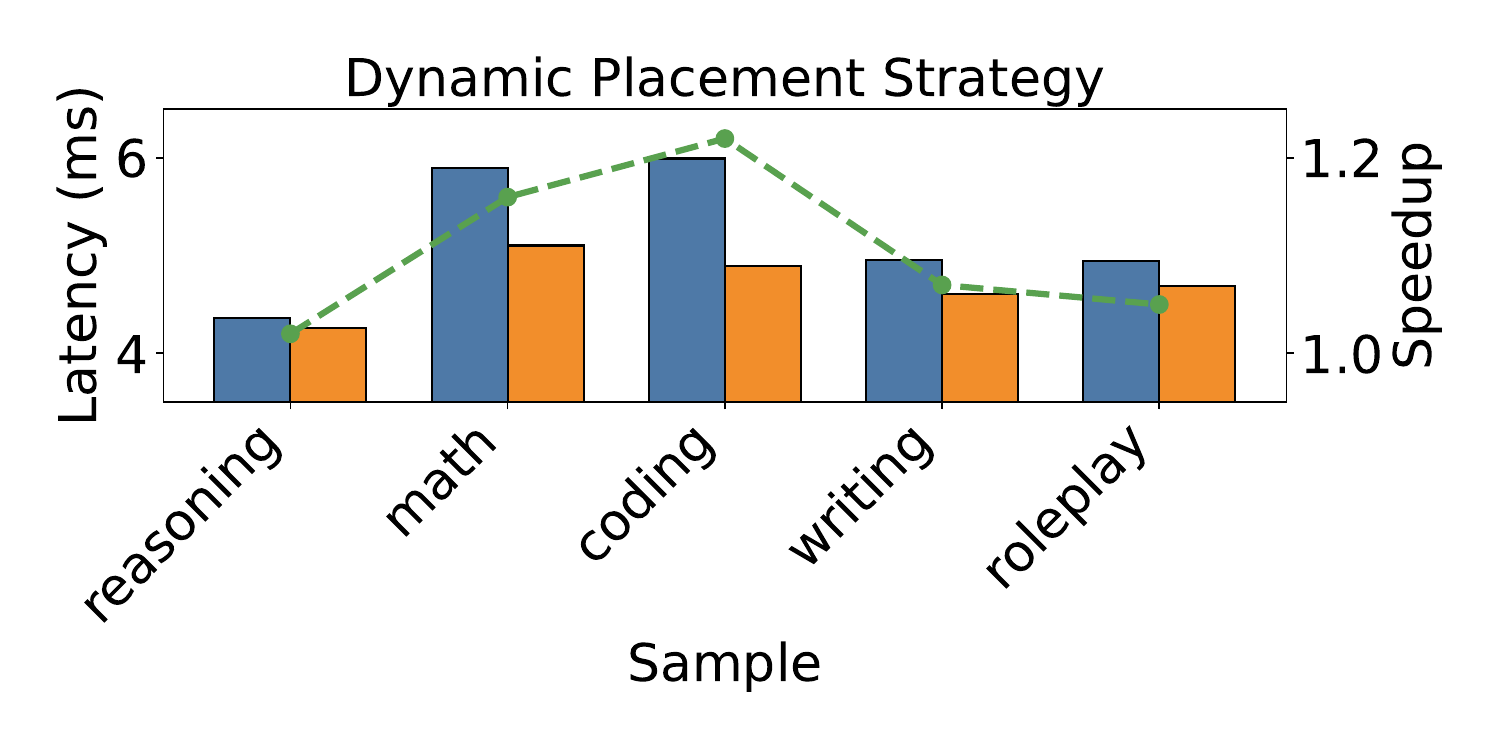}
    \label{fig:dynamic2}
}
\hfill
\subfloat{
    \includegraphics[width=0.45\columnwidth]{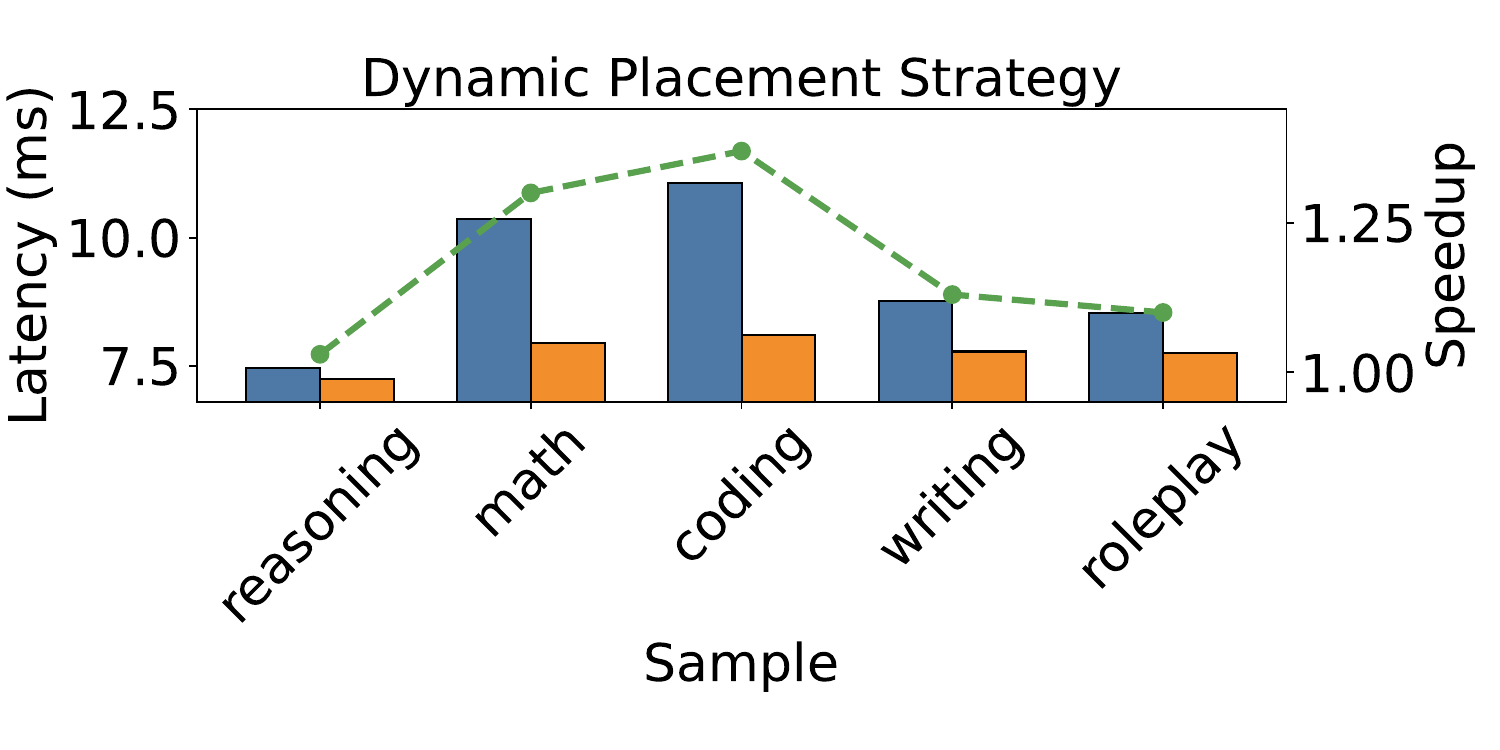}
    \label{fig:dynamic5}
}

\caption{Static vs. dynamic placement. (a) 2 experts pre-broadcast, (b) 5 experts pre-broadcast.}
\label{fig:dynamic-results}
\end{figure}

\subsubsection{Dynamic Placement Strategy}

% To assess the impact of the Dynamic Placement Strategy, we compare the performance of static and dynamic expert placement strategies. In these experiments, we sample multiple expert routing traces in various types of questions from the MT Bench dataset, focusing on tasks with varying expert activation patterns. We compare the latency and speedup between static (generating from reasoning questions) and dynamic strategies under two different hardware configurations and expert broadcasting settings:

% Hardware Configuration: (5 TFLOPS, 50 GB/s bandwidth) with 512 batch size, which has enough time to pre-broadcast 2 experts per layer, the results are shown in Fig. \ref{dynamic}(a).

% Hardware Configuration: (2.5 TFLOPS, 75 GB/s bandwidth) with 512 batch size, which has enough time to pre-broadcast 5 experts per layer, the results are shown in Fig. \ref{dynamic}(b).

We compare static and dynamic expert placement using MT-Bench traces from different categories. Static placement is derived from reasoning traces, while dynamic placement adapts to runtime activation. Fig.~\ref{fig:dynamic-results} reports 5 TFLOPS/50 GB/s with 2 pre-broadcast experts and 2.5 TFLOPS/75 GB/s with 5.

\textbf{Better performance in various scenarios:} 
% The results are shown in Fig. \ref{dynamic}, which indicates that the Dynamic Placement Strategy provides significant speedups and maintains relatively stable inference latency across a variety of real-time inference scenarios. Notably, for tasks such as math and coding problems, which have huge differences from reasoning, the dynamic approach significantly reduces MoE layer latency compared to static deployments. Specifically, when broadcasting 2 experts per layer, the average speedup achieved by the dynamic strategy is 1.15$\times$, and when broadcasting 5 experts per layer, the average speedup increases to 1.25$\times$.
% These results highlight the effectiveness of dynamic expert scheduling in reducing latency by adapting to inference-time workload and improving both computation and communication efficiency.
Dynamic placement reduces latency, especially for math and coding traces that differ from reasoning, and averages \textbf{1.15$\times$} and \textbf{1.25$\times$} speedup with 2 and 5 pre-broadcast experts, respectively.

\subsubsection{Hardware-aware Gating}

We evaluate hardware-aware gating on GSM8K with DeepSeek across batch sizes under 2.5 TFLOPS/75 GB/s. 

\begin{figure}[!tb]
    \centering
    % \scriptsize
    \begin{tabular}{cccc}
        \textcolor{color4}{\rule{1em}{1em}} Static & 
        \textcolor{color1}{\rule{1em}{1em}} Pre & 
        \textcolor{color2}{\rule{1em}{1em}} Adap & 
        \textcolor{color3}{\rule{1em}{1em}} Pre+Adap\\
    \end{tabular} \\[0.1em]
\centerline{\includegraphics[width=\linewidth]{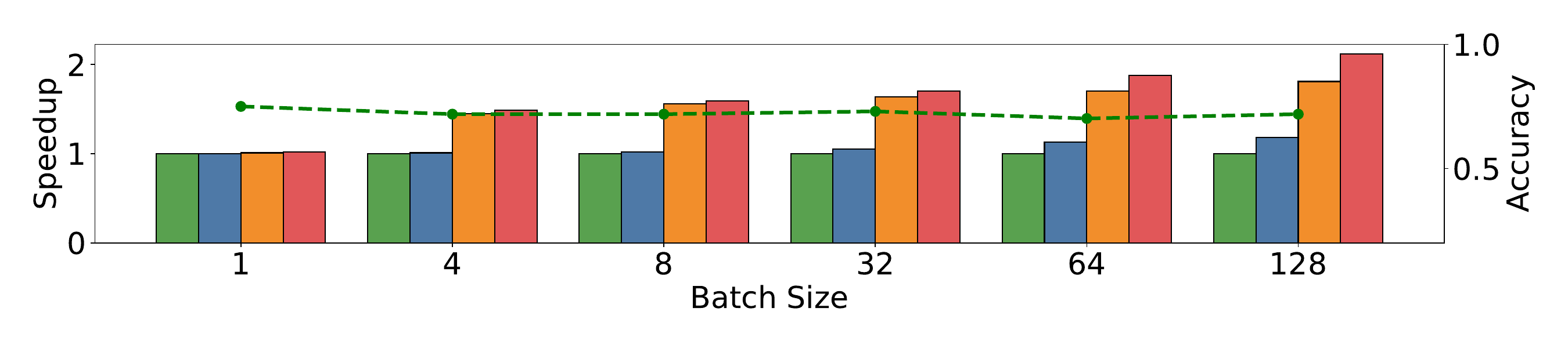}}
\caption{Scalability of Hardware-aware Gating across batch sizes.}
\label{adap}
\end{figure}

\textbf{Better Scalability to Small Batch: }\textcolor{red}{Fig.~\ref{adap} shows that pre-broadcasting weakens at small batches because fewer tokens reveal hot experts, whereas hardware-aware gating remains effective by adjusting low-impact choices with marginal compute/communication penalties. Combining them improves speedup by reducing both hotspot execution cost and the routing demand that creates hotspots.}
% where expert selection occurs before any meaningful compute load has been accumulated, rendering the gating function equivalent to the standard softmax.

Accuracy stays around 70\%, comparable to original gating.

%\subsection{Robustness}

\subsection{\textcolor{red}{Layer-wise Sensitivity of Gating Parameters}}

\begin{figure}[!tb]
    \centering
    \includegraphics[width=\linewidth]{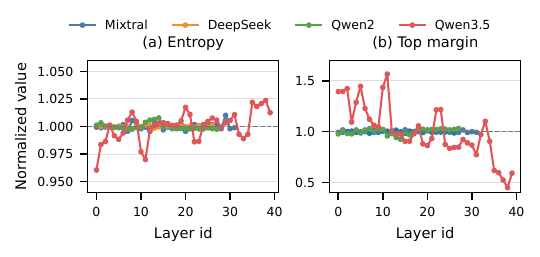}
    \caption{\textcolor{red}{Layer-wise routing statistics normalized by the model-wise mean. Top margin is $s_{(1)}-s_{(k+1)}$.}}
    \label{fig:layerwise-entropy}
\end{figure}

\textcolor{red}{We examine whether hardware-aware gating coefficients should be tuned per layer. Fig.~\ref{fig:layerwise-entropy}(a) shows that normalized gating entropy stays close to 1 across layers, suggesting that the routing-score scale is stable enough to share each model-and-hardware-specific coefficient pair across layers.}

\textcolor{red}{Fig.~\ref{fig:layerwise-entropy}(b) shows larger top-margin variation on Qwen3.5, where top margin is $s_{(1)}-s_{(k+1)}$. However, the top-margin-scaled variant in Table~\ref{tab:layerwise-scale} provides no clear overall benefit: KL and PPL ratio improve slightly, Top-1 agreement is unchanged, and RMS shift increases. Given the added calibration cost, we retain one coefficient pair per model and hardware configuration for all layers.}
\begin{table}[!tb]
\begingroup
\captionsetup{labelfont={color=red},textfont={color=red}}
\color{red}
\centering
\caption{Model-wise fixed vs. layer-wise scaled gating on Qwen3.5.}
\label{tab:layerwise-scale}
\setlength{\tabcolsep}{4pt}
\small
\resizebox{\columnwidth}{!}{
\begin{tabular}{@{}lcccc@{}}
\hline\hline
Setting & KL $\downarrow$ & PPL ratio $\downarrow$ & Top-1 agree. $\uparrow$ & RMS shift $\downarrow$ \\
\hline
Model-wise fixed (final) & 0.0501 & 1.0441 & 91.00\% & 6.77\% \\
Layer-wise scale (diagnostic) & 0.0441 & 1.0341 & 90.98\% & 7.06\% \\
\hline\hline
\end{tabular}}
\endgroup
\end{table}

\subsection{Sensitivity to Gating Parameters}

We analyze how $r_{\text{comp}}$ and $r_{\text{comm}}$ affect MoE-layer accuracy and speedup on DeepSeek under batch size 32.

\begin{figure}[!tb]
\centering
% \scriptsize

\subfloat{
    \includegraphics[width=0.45\columnwidth]{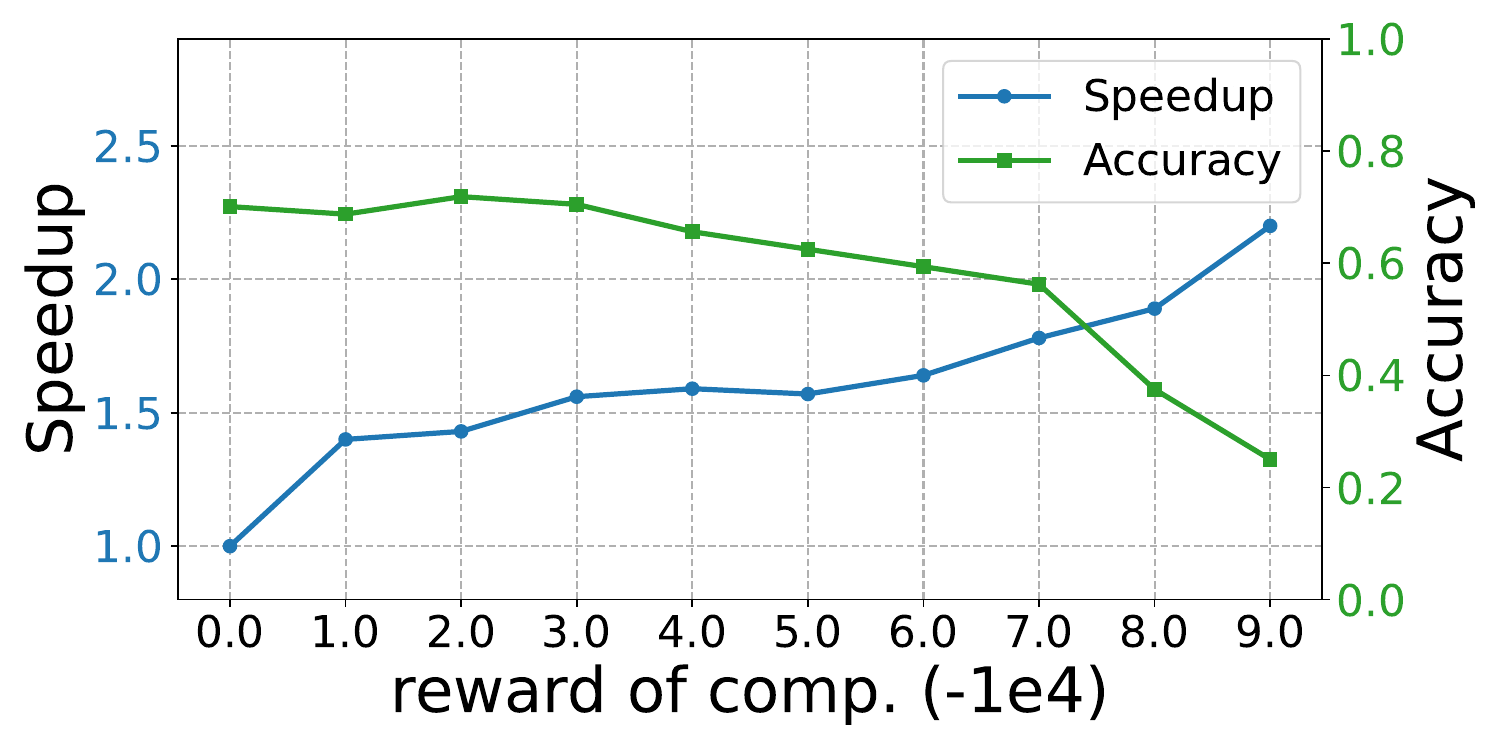}
    \label{adap_comp}
}
\hfill
\subfloat{
    \includegraphics[width=0.45\columnwidth]{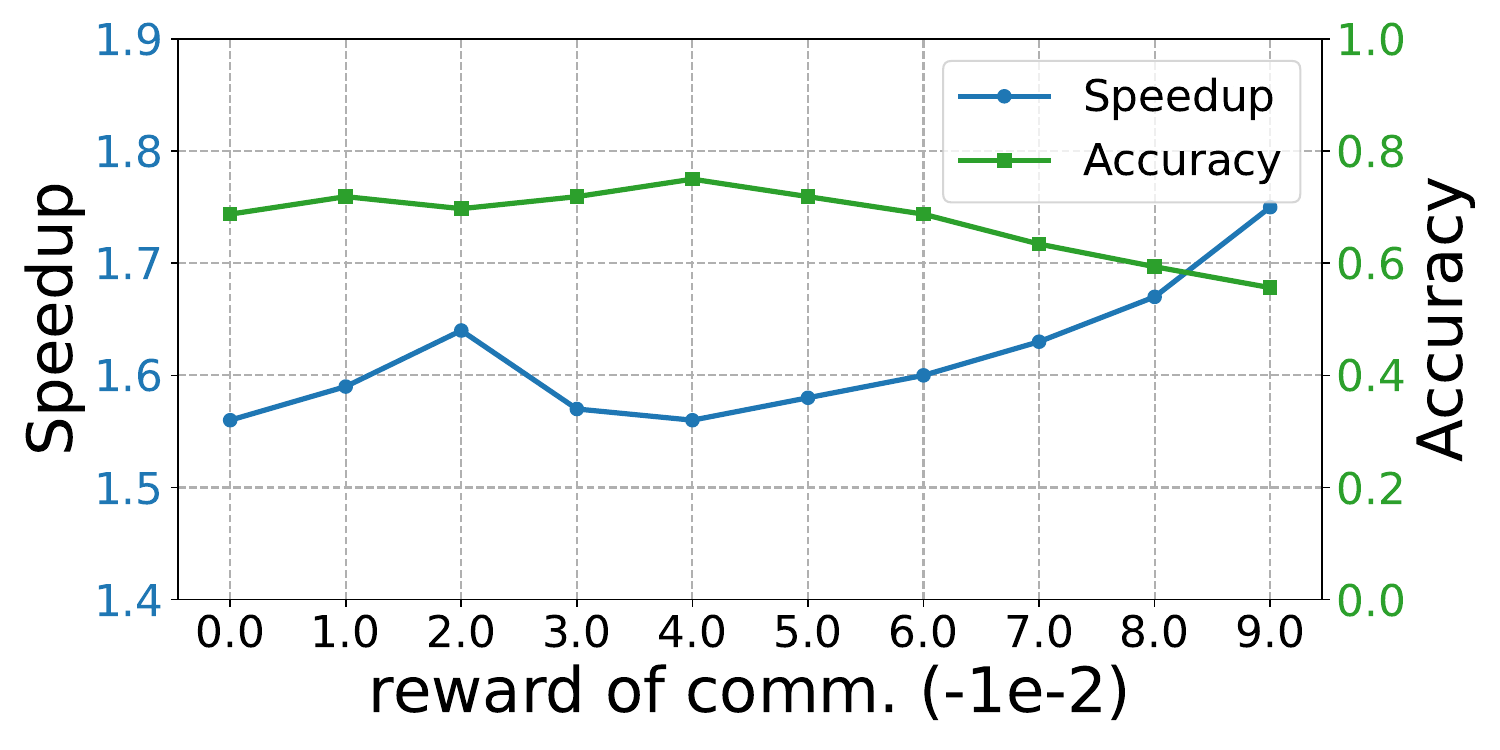}
    \label{adap_comm}
}

\caption{Impact of (a) compute penalty $r_{\text{comp}}$ and (b) communication penalty $r_{\text{comm}}$.}
\label{adap_ablation}
\end{figure}

Fig.~\ref{adap_ablation}(a) shows that increasing $|r_{\text{comp}}|$ favors lighter-compute experts and improves speedup; accuracy stays around 70\% up to $r_{\text{comp}}=-3\times 10^{4}$ and then degrades.

% Fig.~\ref{adap_ablation}(b) evaluates the impact of $r_{\text{comm}}$, which penalizes expert selections that span more nodes. When increasing the magnitude of $r_{\text{comm}}$(also negative), we observe an overall upward trend in speedup, as token dispatch and aggregation involve fewer nodes. However, small fluctuations appear, likely due to increased contention on nodes caused by overly concentrated expert routing. Accuracy remains mostly stable, with a slight drop when $r_{\text{comm}}<-6\times 10^{-2}$. We therefore set $r_{\text{comm}}=-2\times 10^{-2}$ by default to balance performance and stability.

Fig.~\ref{adap_ablation}(b) shows that $r_{\text{comm}}$ penalizes experts with higher deployment dispersion. Speedup is non-monotonic because communication reduction can conflict with compute balance; accuracy stays stable except when $\left |  {r_{\text{comm}}} \right | > 0.06$.

% \begin{figure}[!tb]
% \centering
% \scriptsize

% \begin{subfigure}[t]{0.24\textwidth}
%     \includegraphics[width=\linewidth]{figs/adaptive_performance.pdf}
%     \caption{}
%     \label{adap_comp}
% \end{subfigure}
% \hfill
% \begin{subfigure}[t]{0.24\textwidth}
%     \includegraphics[width=\linewidth]{figs/adaptive_performance_comm.pdf}
%     \caption{}
%     \label{adap_comm}
% \end{subfigure}
% \caption{(a) Impact of Compute Reward (Penalty) $r_{\text{comp}}$ on Accuracy and Speedup, (b) Impact of Communication Reward (Penalty) $r_{\text{comm}}$ on Accuracy and Speedup.}
% \label{adap_ablation}
% \end{figure}

%% file: docs/6-conclusion.tex
\section{Conclusion}

This work presents \method, a framework for MoE inference on 3D NMP that combines hybrid expert placement, runtime scheduling, and routing adaptation. Experiments show that \method~ consistently outperforms existing parallelization strategies, achieving 1.1$\times$--3.4$\times$ over TP, 1.1$\times$--1.5$\times$ over EP, 1.1$\times$--3.7$\times$ over the Hybrid TP-EP baseline, and 1.1$\times$--1.3$\times$ over HD-MoE.